\documentclass[11pt,showpacs,preprintnumbers]{revtex4}
\usepackage{amssymb}
\usepackage{amsmath}
\usepackage{latexsym}
\def\setR{\mathbb{R}}
\def\setN{\mathbb{N}}
\def\setC{\mathbb{C}}
\def\K{\textsf{K}}
\def\1{\mathbf {Id} }
\def\z {{\cal Z}}
\def\y {{\cal Y}^{\nu}({\cal Z)}}
\def\dz {\frac{d}{d{\cal Z}}\,}
\def\ddz {\frac{d^2}{d{\cal Z}^2}\,}
\def\bc{\bar\partial_{\alpha}}
\def\bpc{\bar\partial'_{\beta'}}
\def\ab{\frac{\partial x^{\alpha}}{\partial X^a}\frac{\partial x'^{\beta'}}{\partial X'^{b'}}\;}
\def\a{\frac{\partial x^{\alpha}}{\partial X^a}\;}
\def\bb{\frac{\partial x'^{\beta'}}{\partial X'^{b'}}\;}
\makeatletter
\renewcommand\thesection{\@Roman\c@section}
\renewcommand\theequation{\@arabic\c@section.\@arabic\c@equation}
\makeatother
\begin{document}
%\preprint{...}

\title{``Massive'' spin-$2$ field in de Sitter space}

\author{
T. Garidi$^{1,2}$, J-P.Gazeau$^{1,2}$ and M.V. Takook$^{3,4}$\\
{\it $1$ - LPTMC, Universit\'e Paris 7  Denis Diderot, boite 7020
F-75251 Paris Cedex 05, France.\\
$2$ - F\'ed\'eration de recherche APC, Universit\'e Paris 7
Denis Diderot, boite 7020, \\F-75251 Paris Cedex 05, France.\\
$3$ - Department of Physics, Razi University, Kermanshah, IRAN,\\
$4$ - Plasma Physics research centre, Islamic Azad
University,P.O.BOX 14835-157, Tehran, IRAN }}
\email{garidi@ccr.jussieu.fr, gazeau@ccr.jussieu.fr,
takook@ccr.jussieu.fr}

\date{\today}% It is always \today, today,

\begin{abstract}

In this paper we present a covariant quantization of the
``massive'' spin-$2$ field on de Sitter (dS) space. By ``massive''
we mean a field which carries a specific principal series
representation of the dS group. The work is in the direct
continuation of previous ones concerning the scalar, the spinor
and the vector cases. The quantization procedure,  independent of
the choice of the coordinate system, is based on the
Wightman-G\"arding axiomatic and on analyticity requirements for
the two-point function in the complexified pseudo-Riemanian
manifold. Such a construction is necessary in view of preparing
and comparing with the dS conformal spin-$2$ massless case (dS
linear quantum gravity) which will be considered in a forthcoming
paper and for which specific quantization methods are needed.
\end{abstract}


\maketitle\newpage

%------------------------------------------------------------------------------------------------------------------------
\section{Introduction}

As recent observational data clearly favors a positive
acceleration of the present universe, the de Sitter model
represents an appealing first approximation of the background
space-time. In two previous papers \cite{gata,bagamota},
quantizations of  ``massive" spinor fields and  vector fields on
the dS space have been considered. The spin-$2$ case is of great
importance since the massless tensor field (spin-$2$) is among the
central objects in quantum cosmology and quantum gravity on dS
space (dS linear quantum gravity). It has been found that the
corresponding propagator (in the usual linear approximation for
gravitational field)  exhibits a pathological behaviour for large
separated points (infrared divergence) \cite{altu, flilto,anmo1}.

On one hand this behaviour may originate from the gauge invariance
of the field equation and so should have no physical consequences.
Antoniadis, Iliopoulos and Tomaras \cite{anilto2} have shown that
the large-distance pathological behavior of the graviton
propagator on dS background does not manifest itself in the
quadratic part of the effective action in the one-loop
approximation. This means the pathological behaviour of the
graviton propagator may be gauge dependent and so should not
appear in an effective way as a physical quantity.

On the other hand some authors argue that infrared divergence
could be exploited in order to create instability of dS space
\cite{for, anilto1}. Tsamis and Woodard have considered the field
operator for linear gravity in dS space along the latter line in
terms of flat coordinates, which cover only one-half of the dS
hyperboloid \cite{tswo}. Hence they have found a quantum field
which breaks dS invariance, and  they have examined the resulting
possibility of quantum instability.

Nevertheless, a fully covariant quantization of the linear
gravitational field without infrared divergence in dS space-time
may reveal to be of extreme importance for further developments.
It will be considered in a forthcoming paper \cite{gareta2}. Such
a quantization requires preliminary covariant quantizations of the
minimally coupled scalar field and the ``massive'' spin-$2$ field
respectively.

Recently, de Vega and al. \cite{vera} have shown that,  in flat
coordinates (not global)  on de Sitter space-time, the infrared
divergence does not appear in the ``massless" minimally coupled
scalar field. The question  of the covariant minimally coupled
scalar field has been completely answered in \cite{gareta1} after
introducing a specific Krein QFT. We have shown that the effect of
that quantization, without changing the physical content of the
theory, appears as an automatic renormalization of the ultraviolet
divergence in the stress tensor and of the infrared divergence in
the two-point function \cite{ta4}. By using this method for linear
gravity (the traceless rank-2 ``massless'' tensor field) the
two-point function is free of any infrared divergence \cite{ta}.
This result has been also obtained by \cite{hahetu, hiko1, hiko2}.

Here, we present a fully covariant quantization of the ``massive''
spin-$2$ field. Our method is based on a rigorous
group-theoretical approach combined with a suitable adaptation of
the Wightman-G\"arding axiomatic, which is carried out in terms of
coordinate independent dS waves. The whole procedure originated by
\cite{brgamo} is based on analyticity requirements in the
complexified pseudo-Riemanian manifold. The SO$(1,N)$ unitary
irreducible representations (UIR) acting on symmetric, traceless
and divergence-free tensor eigenfunctions of the Laplace-Beltrami
operator have been investigated in \cite{higu1}. Previous studies
of the ``massive" spin-2 field have been carried out in
\cite{gasp} with a specific choice of coordinate (flat
coordinates) covering only one-half of the dS hyperboloid, and in
\cite{higu2} where the forbidden mass range for spin-2 fields has
been clarified, and the null-mass limit considered. This limit has
also been analyzed in \cite{papa} and recently a consistent theory
for a massive spin-$2$ field in a general gravitational background
has been presented in \cite{bu}.

In section II, we describe the dS tensor field equation as an
eigenvalue equation of the SO$(1,4)$ Casimir operators. The
notations and the two independent Casimir operators are
introduced. It will be convenient to use ambient space notations
in order to express the spin-2 field equation in terms of the
coordinate independent Casimir operators. The latter carry the
group-theoretical content of the theory and it will be reminded
how they enable us to classify the dS group UIR \cite{dix,tak}
according to two parameters $p$ and $q$ which behave like a spin
($s$) and a mass ($m$) in the Minkowskian limit, depending on the
nature of the involved group representation.

Section III is devoted to the field equation and its solutions.
The dS tensor modes are written in terms of a scalar field $\phi$
and a generalized polarization tensor ${\cal E}$ $$ \K_{\alpha
\beta} (x)= {\cal E}_{\alpha \beta}(x,\xi)\phi (x). $$ As for
spinor and vector fields,  the tensor ${\cal E}(x,\xi)$ is a
space-time function in dS space-time. There is a certain extent of
arbitrariness in the choice of this tensor  and we fix it in such
a way that, in the limit $H=0$, one obtains the polarization
tensor in Minkowski space-time.

In section IV we derive the Wigthman two-point function ${\cal
W}_{\alpha\beta \alpha'\beta'}(x,x')$. This function fulfills the
conditions of : a) positiveness, b) locality, c) covariance, d)
normal analyticity, e) transversality, f) divergencelessness and
g) permutational index symmetries. The four conditions c), e), f),
and g) allow one to associate this field with a spin-$2$ unitary
irreducible representation of the dS group. The positivity
condition permits us to construct a Hilbert space structure. The
locality is related to the causality principle, which is a well
defined concept in dS space. The normal analyticity allows one to
view ${\cal W}_{\alpha\beta \alpha'\beta'}(x,x')$ as the boundary
value of an analytic two-point function $W_{\alpha\beta
\alpha'\beta'}(z,z')$ from the tube domains. The analytic kernel
$W_{\alpha\beta \alpha'\beta'}(z,z')$ is defined in terms of dS
waves in their tubular domains. Then, the Hilbert space structure
is made explicit and the field operator ${\cal K} (f)$ is derived.
We also give a coordinate-independent formula for the unsmeared
field operator ${\cal K}(x)$. Brief conclusion and outlook are
given in section V. It is in particular asserted that the
extension of our approach to ``massless'' tensor field (
gravitational field in a dS background in the linear
approximation) requires an indecomposable representation of the dS
group in view of the construction of the corresponding covariant
quantum field. Finally, we have detailed the classification of the
unitary representation of SO$_{0}(1,4)$ in appendix A. In appendix
B we relate our construction to the maximally symmetric bitensors
introduced in Reference \cite{allen}. In appendix C and D we
respectively present the ``massive'' vector and tensor two-point
functions.
%-------------------------------------------------------------------------------------------------------------------------
\section{Field equations on de Sitter space}
\subsection{Ambient space notations and Casimir operators}
The de Sitter space is a solution of the cosmological Einstein
equation with positive cosmological constant $\Lambda$. It is
conveniently described as a hyperboloid embedded in a
five-dimensional Minkowski space
\begin{equation}
X_H=\{x \in \setR^5 ;x^2=\eta_{\alpha\beta} x^\alpha x^\beta
=-H^{-2}=-\frac{3}{\Lambda}\},\;\;\alpha,\beta=0,1,2,3,4,
\end{equation}
where $\eta_{\alpha\beta}=$ diag$(1,-1,-1,-1,-1)$. The de Sitter
metrics reads
$$ds^2=\eta_{\alpha\beta}dx^{\alpha}dx^{\beta}=g_{\mu\nu}^{dS}dX^{\mu}dX^{\nu},\;\;\mu=0,1,2,3,$$
where the $X^\mu$'s are  $4$ space-time intrinsic coordinates of
the dS hyperboloid.

An immediate  realization space is made of a second-rank intrinsic
tensor field $h_{\mu\nu}$ satisfying the conditions of
divergenceless, tracelessness, and index permutational symmetry
respectively:
\begin{equation}
\nabla.h(X)=0,\;\;h_{\mu}^{\mu}(X)=0,\;\;h_{\mu\nu}=h_{\nu\mu}.
\label{eq:con1}
\end{equation}
The wave equation for such fields propagating in de Sitter space
can be written as \cite{gasp}
\begin{equation}\label{eq:wav}
\left(\Box_H +2H^2+m_H^2\right) h_{\mu\nu}(X)=0,
\end{equation}
where $\Box_H=\nabla_{\mu}\nabla^{\mu}$ is the d'Alembertian
operator.

Let us now adopt ambient space notations (for details see
\cite{fr}), namely ${\cal K}_{\alpha\beta}(x)$ for the field. With
these notations, the relationship with unitary irreducible
representations of the dS group becomes straightforward  because
the Casimir operators are easy to identify. The tensor field
${\cal K}_{\alpha\beta}(x)$ has to be viewed as a homogeneous
function of the $\setR^5$-variables $x^{\alpha}$ with homogeneous
degree $\lambda$ and thus satisfies ,
\begin{equation}
x^{\alpha}\frac{\partial }{\partial x^{\alpha}}{\cal
K}_{\gamma\beta}(x)= x.\partial {\cal
K}_{\gamma\beta}(x)=\lambda\, {\cal K}_{\gamma\beta}(x).
\end{equation}
The direction of ${\cal K}_{\alpha\beta}(x)$  lies in the de
Sitter space if we require the condition of transversality
\cite{di}
\begin{equation}
x.{\cal K}(x)=0. \label{eq:tra}
\end{equation}
With these notations, the conditions $(\ref{eq:con1})$ read as
\begin{equation}
\bar \partial. {\cal K}=0,\;\; {\cal K}_{\alpha}^{\alpha}={\cal
K}'=0,\;\; {\cal K}_{\alpha\beta}={\cal K}_{\beta\alpha},
\label{eq:con2}
\end{equation}
where $\bar\partial$ is the tangential (or transverse) derivative
on dS space,
\begin{equation}
\bar\partial_\alpha=\theta_{\alpha\beta}\partial^\beta=\partial_\alpha
+H^2x_\alpha x.\partial,\quad\mbox{with}\quad x.\bar\partial=0.
\end{equation}
The tensor with components
$\theta_{\alpha\beta}=\eta_{\alpha\beta}+H^2x_{\alpha}x_{ \beta}$
is the so-called transverse projector.

In order to express Equation (\ref{eq:wav}) in terms of the
ambient coordinates, we use the fact that the ``intrinsic'' field
$h_{\mu\nu}(X)$ is locally determined by the transverse tensor
field ${\cal K}_{\alpha\beta}(x)$ through
\begin{equation}
h_{\mu\nu}(X)=\frac{\partial x^\alpha}{\partial X^\mu}
\frac{\partial x^\beta} {\partial X^\nu}{\cal
K}_{\alpha\beta}(x(X)).
\end{equation}
For instance, it is easily shown that the metric $\eta_{\mu\nu}$
corresponds  to the transverse projector $\theta_{\alpha\beta}$.
Covariant derivatives acting on a l-rank tensor are transformed
according to
\begin{equation}
\nabla_{\mu}\nabla_{\nu}..\nabla_{\rho}h_{\lambda_{1}..\lambda_{l}}=
\frac{\partial x^\alpha}{\partial X^\mu} \frac{\partial x^\beta}
{\partial X^\nu}..\frac{\partial x^\gamma}{\partial X^\rho}
\frac{\partial x^{\eta_{1}}}{\partial X^{\lambda_{1}}}
..\frac{\partial x^{\eta_{l}}} {\partial X^{\lambda_{l}}}
\mbox{Trpr}\bar{\partial}_{\alpha}\mbox{Trpr}\bar{\partial}_{\beta}
..\mbox{Trpr}\bar{\partial}_{\gamma}{\cal
K}_{\eta_{1}..\eta_{l}}\,,
\end{equation}
where the transverse projection defined by
\begin{equation} \left(\mbox{Trpr}
\,{\cal
K}\right)_{\lambda_{1}..\lambda_{l}}\equiv\theta^{\eta_{1}}_{\lambda_{1}}
..\theta^{\eta_{l}}_{\lambda_{l}}{\cal
K}_{\eta_{1}..\eta_{l}}\,,\nonumber
\end{equation}
guarantees  the transversality  in each index. Applying this
procedure to a transverse second rank, symmetric tensor field,
leads to
\begin{eqnarray}\label{eqn:cova}
\nabla_{\mu}\nabla_{\nu}h_{\rho\lambda}&=& \frac{\partial
x^\alpha}{\partial X^\mu} \frac{\partial x^\beta} {\partial
X^\nu}\frac{\partial x^\gamma}{\partial X^\rho} \frac{\partial
x^{\eta}}{\partial X^{\lambda}}
\mbox{Trpr}\bar{\partial}_{\alpha}\mbox{Trpr}
\bar{\partial}_{\beta}{\cal K}_{\gamma\eta} \nonumber\\
&=&\frac{\partial x^\alpha}{\partial X^\mu} \frac{\partial
x^\beta} {\partial X^\nu}\frac{\partial x^\gamma}{\partial X^\rho}
\frac{\partial x^{\eta}}{\partial X^{\lambda}}
\left(\bar{\partial}_{\alpha}\bar{\partial}_{\beta}{\cal
K}_{\gamma\eta}-H^{2} \theta_{\alpha\gamma}{\cal
K}_{\beta\eta}-H^{2}\theta_{\alpha\eta}{\cal K}_{\beta\gamma}
\right)\,.
\end{eqnarray}

The kinematical group of the de Sitter space is the $10$-parameter
group SO$_0(1,4)$ (connected component of the identity in
SO$(1,4)$ ), which is one of the two possible deformations of the
Poincar\'e group. There are two Casimir operators
\begin{equation}
Q^{(1)}_2=-\frac{1}{2}L_{\alpha\beta}L^{\alpha\beta},\qquad
Q^{(2)}_2=-W_{\alpha}W^{\alpha},\label{eq:cas}
\end{equation}
where
\begin{equation}
W_{\alpha}=-\frac{1}{8}\epsilon_
{\alpha\beta\gamma\delta\eta}L^{\beta\gamma}L^{\delta\eta},
\quad\mbox{with  10 infinitesimal generators}\quad
L_{\alpha\beta}=M_{\alpha\beta}+S_{\alpha\beta}.
\end{equation}
The subscript $2$ in $Q^{(1)}_2$, $Q^{(2)}_2$ reminds that the
carrier space is constituted by  second rank tensors. The orbital
part $M_{\alpha\beta}$, and the action of the spinorial part
$S_{\alpha\beta}$ on a tensor field ${\cal K}$ defined on the
ambient space read respectively \cite{gaha}
$$M_{\alpha\beta}=-i (x_\alpha\partial_\beta-x_\beta\partial_\alpha),$$
\begin{equation}
S_{\alpha\beta}{\cal K}_{\gamma\delta}=-i(\eta_{\alpha\gamma}{\cal
K}_{\beta\delta}-\eta_{\beta\gamma} {\cal
K}_{\alpha\delta}+\eta_{\alpha\delta}{\cal
K}_{\beta\gamma}-\eta_{\beta\delta}{\cal K}_{\alpha\gamma}).
\label{eq:spi}
\end{equation}
The symbol $\epsilon_{\alpha\beta\gamma\delta\eta}$ holds for the
usual antisymmetrical tensor. The action of the Casimir operator
$Q_2^{(1)}$  on $\K$ can be written in the more explicit form
\begin{equation}\label{eq:act}
Q_2^{(1)}{\cal K}(x)=\left(Q_{0}^{(1)}-6\right){\cal K}(x)+2\eta
{\cal K}'+2{\cal S} x\partial\cdot{\cal K}(x)-2{\cal S}  \partial
x\cdot{\cal K}(x),
\end{equation}
In the latter,
$Q_{0}^{(1)}=-{{1}\over{2}}M_{\alpha\beta}M^{\alpha\beta}$, and
the vector symmetrizer ${\cal S}$ is defined for two vectors
$\xi_{\alpha}$ and $\omega_{\beta}$ by ${\cal
S}(\xi_{\alpha}\omega_{\beta})=\xi_{\alpha}\omega_{\beta}+\xi_{\beta}\omega_{\alpha}$.

We are now in position to express the wave equation
$(\ref{eq:wav})$ by using the Casimir operators. This can be done
with the help of Equation (\ref{eqn:cova}) since
$Q_{0}^{(1)}=-H^{-2}(\bar\partial)^2$. The d'Alembertian operator
becomes
\begin{equation}
\Box_{H}
h_{\mu\nu}=\nabla^{\lambda}\nabla_{\lambda}h_{\mu\nu}=-\frac{\partial
x^\alpha}{\partial X^\mu} \frac{\partial x^\beta}{\partial X^\nu}
\left[Q_{0}^{(1)}H^2+2H^2\right]{\cal K}_{\alpha\beta} \;,
\end{equation}
and the wave equation $(\ref{eq:wav})$ is rewritten as
\begin{equation}
\left(Q_{0}^{(1)}-H^{-2}m_H^2\right){\cal K}_{\alpha\beta}(x)=0.
\end{equation}
Finally, using formula (\ref{eq:act}) for the tensor field ${\cal
K}_{\alpha\beta}(x)$ which satisfies the conditions
$(\ref{eq:con2})$ the field equation becomes
\begin{equation}
\left( Q_{2}^{(1)}-\left( m^{2}_{H}H^{-2}-6\right) \right){\cal
K}_{\alpha\beta}(x)=0. \label{eq:wave}
\end{equation}

As expected, this formulation of the field equation has now a
clear group-theoretical content. In fact, using the representation
classification given by the eigenvalues of the Casimir operator,
we will be able to identify the involved field. At this point let
us clarify what we mean by ``massive" spin-2 de Sitter field.
Inasmuch as mass and  spin are well-defined Poincar\'e concepts,
we will consider exclusively  the de Sitter elementary systems (in
the Wigner sense) associated to a UIR of SO$_{0}(1,4)$ that admit
a non-ambiguous  massive spin-2 UIR of the Poincar\'e group at the
$H=0$ contraction limit. This contraction is performed with
respect to the  subgroup SO$_0(1,3)$ which is identified as the
Lorentz subgroup in both relativities, and the concerned de Sitter
representations are precisely those ones which are induced by the
{\it minimal parabolic} \cite{lipsman} subgroup SO$(3) \times $SO
$(1,1) \times$(a certain nilpotent subgroup), where SO$(3)$
\underline{is} the space rotation subgroup of the Lorentz subgroup
in both cases. This fully clarifies the concept of spin in de
Sitter since it is issued from the {\it same} SO$(3)$.
\subsection{``Massive'' spin-2 unitary representation of the de Sitter group {SO}$_{0}(1,4)$}
The operator $Q_2^{(1)}$ commutes with the action of the group
generators and, as a consequence, it is constant in each unitary
irreducible representation (UIR). Thus the eigenvalues of
$Q_2^{(1)}$ can be used to classify the UIR's {\it i.e.,}
\begin{equation}
(Q_2^{(1)}-\langle Q_2^{(1)}\rangle){\cal K}(x)=0.
\end{equation}
Following Dixmier \cite{dix} we get a classification scheme using
a pair $(p,q)$ of parameters involved in the following possible
spectral values of the Casimir operators :
\begin{equation}
Q^{(1)}=\left(-p(p+1)-(q+1)(q-2)\right)I_d ,\qquad\quad
Q^{(2)}=\left(-p(p+1)q(q-1)\right)I_d\,.
\end{equation}
Three types of scalar, tensorial or spinorial UIR are
distinguished for SO$_{0}(1,4)$ according to the range of values
of the parameters $q$ and $p$ \cite{dix,tak}, namely : the
principal, the complementary and the discrete series. In the
following, we shall restrict  the list to the unitary
representations which have a Minkowskian physical spin-2
interpretation in the limit $H=0$ (for the general situation see
\cite{babo} and Appendix A). The flat limit tells us that for the
principal and the complementary series it is the value of $p$
which has a spin meaning, and that, in the case of the discrete
series, the only representations which have a physically
meaningful Minkowskian counterpart are those with $p=q$ (details
about the mathematics of the group contraction and the physical
principles underlying the relationship between de Sitter and
Poincar\'e groups can be found in \cite{nah} and \cite{lev}
respectively). The spin-$2$ tensor representations relevant to the
present work are the following :
\begin{itemize}
\item[i)] The UIR's $U^{2,\nu}$ in the principal series where
$p=s=2$ and $q=\frac{1}{2 }+i\nu$ correspond to the Casimir
spectral values:
\begin{equation}
\langle Q_2^{(1)}\rangle=\nu^2-\frac{15}{4},
\end{equation}
with parameter
$\nu \in \setR$ (note that $U^{2,\nu}$ and $U^{2,-\nu}$ are
equivalent).
\item[ ii)] The UIR's $V^{2,q}$ in the complementary series where
$p=s=2$ and $q-q^2=\mu,$ correspond to
\begin{equation}
\langle  Q_2^{(1)}\rangle=q-q^2-4\equiv \mu-4,\;\;\;0<\mu<\frac{1}{4}\,.
\end{equation}
\item[iii)] The UIR's $\Pi^{\pm}_{2,2}$ in the discrete series where
$q=p=s=2$ correspond to
\begin{equation}
\langle Q_2^{(1)}\rangle=-6 .
\end{equation}
The spin-2 ``massless'' field in de Sitter space corresponds to
the latter case in which the sign $\pm$ in $\Pi^{\pm}_{2,2}$
stands for the helicity. A forthcoming paper will be entirely
devoted to this specific field.
\end{itemize}
\noindent Equation $(\ref{eq:wave})$ leads to $H^2(\langle
Q_2^{(1)}\rangle+ \, 6)=m^2_H$ which enables us to write the
respective ``mass'' relations for the three types of UIR
previously described :
\begin{equation}m_{H}^{2}=\;\;
\left \{\begin{array}{rcllr} m^2_p&=&H^2(\nu^2+{\displaystyle
\frac{9}{4}}),\;\;\;\nu\geq 0\;\;\mbox{(for the
principal series),}\vspace{0.3cm} \\
m^2_c&=&H^2(\mu+2),\;\;\;0<\mu<\frac{1}{4}\;\;\mbox{(for the
complementary series),}\vspace{0.3cm} \\
m^2_d&=&0\;\;\mbox{(for the discrete series).}
\end{array}
\right.
\end{equation}
The  spin-$2$ ``mass'' range can be represented by \vspace{0.5cm}
\begin{figure}[h]
\begin{center}
\begin{picture}(0,0)%
\includegraphics{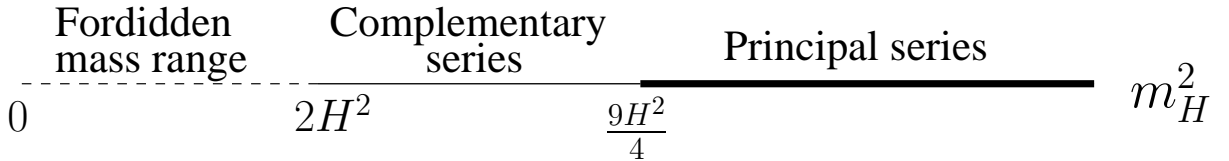}%
\end{picture}%
\setlength{\unitlength}{3947sp}%
\begingroup\makeatletter\ifx\SetFigFont\undefined
% extract first six characters in \fmtname
\def\x#1#2#3#4#5#6#7\relax{\def\x{#1#2#3#4#5#6}}%
\expandafter\x\fmtname xxxxxx\relax \def\y{splain}%
\ifx\x\y   % LaTeX or SliTeX?
\gdef\SetFigFont#1#2#3{%
\ifnum #1<17\tiny\else \ifnum #1<20\small\else \ifnum
#1<24\normalsize\else \ifnum #1<29\large\else \ifnum
#1<34\Large\else \ifnum #1<41\LARGE\else \huge\fi\fi\fi\fi\fi\fi
\csname #3\endcsname}%
\else \gdef\SetFigFont#1#2#3{\begingroup \count@#1\relax \ifnum
25<\count@\count@25\fi
\def\x{\endgroup\@setsize\SetFigFont{#2pt}}%
\expandafter\x \csname \romannumeral\the\count@
pt\expandafter\endcsname \csname @\romannumeral\the\count@
pt\endcsname
\csname #3\endcsname}%
\fi \fi\endgroup
\begin{picture}(7050,879)(1201,-5161)
\put(8251,-4936){\makebox(0,0)[lb]{\smash{\SetFigFont{20}{24.0}{rm}$m_{H}^{2}$}}}
\put(3001,-5086){\makebox(0,0)[lb]{\smash{\SetFigFont{17}{20.4}{rm}$2H^{2}$}}}
\put(1201,-5086){\makebox(0,0)[lb]{\smash{\SetFigFont{17}{20.4}{rm}$0$}}}
\put(4951,-5161){\makebox(0,0)[lb]{\smash{\SetFigFont{17}{20.4}{rm}$\frac{9H^{2}}{4}$}}}
\end{picture}
\caption{Mass range and spin-$2$ SO$_{0}(1,4)$ unitary irreducible
representations.}
\end{center}
\end{figure}

%{Ce qui suit fonctionne aussi  mais pour envoyer sur le web ils demandent du eps}
%\begin{figure}[h]
%\begin{center}
%\input{yguchi.pstex_t}
%\caption{Mass range and spin-$2$ SO$_{0}(1,4)$ unitary irreducible
%representations.}
%\end{center}
%\end{figure}

The forbidden mass range has been discussed by Higuchi in
\cite{higu2} and contrary to his point of view we do not consider
$m_H$ in the range of  the complementary series as a ``mass''.
This is because the complementary series with $p=2$  is not linked
to any physical representation in the Poincar\'e flat limit sense.
The crucial point is that $m^{2}_{c}$ (unlike $m^{2}_{p}$ !) is
confined between the values $0$ and $1/4$ and therefore simply
vanishes in the limit $H=0$. On the contrary, for the principal
series, the contraction limit has to be understood through the
constraint $m= H\nu$. The quantity $m_H$, supposed to depend on
$H$, goes to the Minkowskian mass $m$ when the curvature goes to
zero. In short, we only consider as ``massive'' tensor fields,
those ones for which the values assumed by the parameter $m_H$ are
in the range $m_p$ which corresponds to the principal series of
representations. Eq. $(\ref{eq:wave})$ then gives
\begin{equation}
\left(\Box_H+2H^2+m^2_p\right){\cal K}_{\alpha\beta}(x)=0.
\end{equation}

Let us recall at this point the physical content of the principal
series representation from the point of view of a Minkowskian
observer (at the limit $H=0$). The principal series UIR
$U^{2,\nu},\;\;\nu \geq 0$, contracts toward the tensor massive
Poincar\'e UIR's $P^<(m,2)$ and $P^>(m,2)$ with negative and
positive energies respectively. Actually, the group representation
contraction procedure is not unique and it has been shown that the
principal series UIR can contract either toward the direct sum of
the two tensor massive Poincar\'e UIR's \cite{mini}
\begin{equation}
U^{2,\nu} {{H\rightarrow 0,\nu \rightarrow\infty} \over
H\nu=m}\rightarrow P^<(m,2)\oplus P^>(m,2),
\end{equation}
or simply (forthcoming paper)
\begin{equation} U^{2,\nu}
{{H\rightarrow 0,\nu \rightarrow\infty} \over H\nu=m}\rightarrow
 P^>(m,2)\qquad\mbox{or}\qquad U^{2,-\nu} {{H\rightarrow 0,\nu
\rightarrow\infty} \over H\nu=m}\rightarrow P^<(m,2).
\end{equation}
In contrast, in the massless spin-$2$ case, only the two
aforementioned representations $\Pi^{\pm}_{2,2}$, in the discrete
series with $p=q=2$, have a Minkowskian interpretation. The
representation $\Pi^+_{2,2}$ has a unique extension to a direct
sum of two UIR's $C(3;2,0)$ and $C(-3;2,0)$ of the conformal group
$SO_0(2,4)$ with positive and negative energies respectively
\cite{babo,anla}. The latter restricts to the tensor massless
Poincar\'e UIR's $P^>(0, 2)$ and $P^<(0,2)$ with positive and
negative energies respectively. The following diagrams illustrate
these connections
\begin{equation}
\left.
\begin{array}{ccccccc}
&& {\cal C}(3,2,0)& &{\cal C}(3,2,0)&\hookleftarrow &{\cal P}^{>}(0,2)\\
\Pi^+_{2,2} &\hookrightarrow &\oplus&\stackrel{H=0}{\longrightarrow} & \oplus  & &\oplus\\
&& {\cal C}(-3,2,0)& & {\cal C}(-3,2,0) &\hookleftarrow &{\cal
P}^{<}(0,2),\\
\end{array}
\right.
\end{equation}

\begin{equation}
\left.
\begin{array}{ccccccc}
&& {\cal C}(3,0,2)& &{\cal C}(3,0,2)&\hookleftarrow &{\cal P}^{>}(0,-2)\\
\Pi^-_{2,2} &\hookrightarrow &\oplus&\stackrel{H=0}{\longrightarrow}&\oplus &&\oplus\\
&& {\cal C}(-3,0,2)&& {\cal C}(-3,0,2)&\hookleftarrow &{\cal P}^{<}(0,-2),\\
\end{array}
\right.
\end{equation}

where the arrows $\hookrightarrow $ designate unique extension,
and $ {\cal P}^{ \stackrel{>} {<}}(0,2)$ (resp. $ {\cal
P}^{\stackrel{>}{<}}(0,-2)$) are the massless Poincar\'e UIR's
with positive and negative energies and  positive (resp. negative)
helicity.
%---------------------------------------------------------------------------------------------------------------------
\setcounter{equation}{0}
\section{De Sitter tensor  waves}
\subsection{Field Equation solution}
Our aim is now to solve the ``massive'' spin-2 wave equation for
the dS  mode $\K(x)$
\begin{equation}\label{eq:wave1}
\left(Q_2^{(1)}-\langle
Q_2^{(1)}\rangle\right)\K(x)=0\qquad\mbox{with}\qquad \langle
Q_2^{(1)}\rangle=\nu^{2}-{{15}\over{4}}\,.
\end{equation}
In ambient space  notations, the most general transverse,
symmetric field $\K_{\alpha\beta}(x)$ can be written in terms of
two vector fields $K,K_g$ and a scalar field $\phi$ through the
following recurrence formula \cite{gaha}
\begin{equation}
\K=\theta\phi+{\cal S}\bar Z_1K+D_2K_g,
\label{eq:an}
\end{equation}
with $\K$ satisfying the conditions $(\ref{eq:con2})$. The symbol
$Z_1$ denotes a constant vector and  $\bar
Z_{1\alpha}=\theta_{\alpha\beta}Z_1^\beta\;,x.\bar Z_1=0$. The
operator $D_2$ is the generalized gradient $ D_2K=H^{-2}{\cal
S}(\bar
\partial-H^2 x)K$ which makes a symmetric transverse tensor field
from the transverse vector $K$. The algebraic machinery valid for
describing fields in anti-de Sitter space can be easily
transferred {\it mutatis mutendis} to dS space formalism by the
substitutions (see for instance \cite{gaha,fr,gazeau}):
$$Q_s^{AdS}\longrightarrow -Q_s^{dS},  \;\;\;(H^2)^{AdS}\longrightarrow
-(H^2)^{dS}\, .$$ Reference \cite{gazeau} provides the following
useful relations
\begin{equation}\label{eq:prop4}
\qquad Q_2\theta \phi=\theta Q_0\phi,\qquad
Q_2D_2K_{g}=D_2Q_1K_{g},\nonumber
\end{equation}
\begin{equation}
Q_2{\cal S}\bar Z_1 K={\cal S}\bar Z_1(Q_1-4)K-2H^2D_2(x\cdot
Z_1)K+4\theta (Z_1\cdot K)\,.
\end{equation}
Defining the generalized divergence $\partial_2\cdot \K=\partial
\cdot\K- H^2 x \K'-\frac{1}{2} \bar  \partial \K'$ and
$D_1=H^{-2}\bar \partial$, one also has
\begin{equation}
\partial_2 \cdot
\theta\phi=-H^2D_1\phi,\qquad\partial_2 \cdot D_2K_g =
-(Q_1+6)K_g\,,\nonumber
\end{equation}
\begin{equation}
\partial_2 \cdot {\cal S}\bar Z_1K=\bar Z_1
\partial \cdot K-H^2D_1(Z_{1}\cdot K)-H^2x(Z_{1}\cdot
K)+Z_{1}\cdot\bar\partial K+5H^2(Z_{1}\cdot x)
K.\label{eq:divergenceless}
\end{equation}
 Putting $\K_{\alpha\beta}(x)$ given by
$(\ref{eq:an})$ into $(\ref{eq:wave1})$ and from the linear
independence of the terms in ($\ref{eq:an}$) one gets
\begin{eqnarray}
&&\left(Q_{1}-\langle Q_1^{(1)}\rangle\right)K=0
\quad\mbox{with}\quad\langle Q_1^{(1)}\rangle= \langle Q_2^{(1)}\rangle+4\;,\label{eqn:eqa} \\
&&\left(Q_{0}-\langle Q_2^{(1)}\rangle \right)\phi =-4(Z_1 \cdot K)\;,\label{eqn:eqa1}\\
&&\left(Q_1 -\langle Q_2^{(1)}\rangle  \right)K_g=2H^2 (x \cdot
Z_1) K\label{eqn:eqa2}\;.
\end{eqnarray}
Note that in these formulas, $\langle Q_{s}^{(1)}\rangle$
corresponds to the principal series of representation with spin
$s$ and that $K$ is chosen to be divergenceless. Using the
equations  (\ref{eq:divergenceless}), the divergenceless condition
combined with Eq. (\ref{eqn:eqa2}) leads to
\begin{equation}\label{eq:lek}
K_g=\frac{1}{ \langle Q_{0}^{(1)}\rangle}
\left[-H^{2}D_{1}(\phi+Z_1\cdot K)+ Z_1\cdot\bar\partial
K-H^2xZ_1\cdot K+3H^2 x\cdot Z_1 K \right],
\end{equation}
where  $\langle Q_{0}^{(1)}\rangle=\langle Q_{2}^{(1)}\rangle+6$.
Finally, the traceless condition which yields
\begin{equation}
\bar{\partial}\cdot K_{g}=-2H^{2}\phi -H^{2}Z_1\cdot K\,,
\end{equation}
compared to the divergence of Eq. (\ref{eq:lek}) allows to express
$\phi$ in terms of $K$:
\begin{equation}
\phi=-\frac{2}{3}(Z_1\cdot K). \label{eq:phi}
\end{equation}

Thus, the fields $K$ and $\phi$ are respectively ``massive''
vector field ({\it e.g.} transforming under the vector UIR
$U^{1,\nu}$ of the principal series) \cite{gata}, and ``massive''
scalar field ({\it e.g.} transforming under the scalar UIR
$U^{0,\nu}$ of the principal series) \cite{brgamo}:
\begin{eqnarray}
\left(Q_{1}-\langle
Q_1^{(1)}\rangle\right)K=0\;,\qquad\mbox{and}\qquad
\left(Q_{0}-\langle Q_0^{(1)}\rangle \right)\phi =0\;.
\end{eqnarray}

Note that the equations for $K$ and $\phi$ are compatible with the
relation $\phi=-{{2}\over{3}}Z_{1}\cdot K$. The equations
(\ref{eq:lek}) and (\ref{eq:phi}) show that the massive vector $K$
determines completely the tensor field $\K$ which can now be
written
\begin{equation}\label{eq:wave2}
\K(x)=\left(-\frac{2}{3}\theta Z_1\cdot+{\cal S} \bar
Z_1+\frac{1}{ \langle Q_0^{(1)}\rangle}D_2[Z_1\cdot
\bar\partial-H^2xZ_1\cdot+3H^2 x\cdot Z_1 -\frac{1}{3}H^2 D_1
Z_1\cdot]\right) \,K\,.
\end{equation}

As explained in \cite{gata} the solutions  to Eq.
$(\ref{eqn:eqa})$ are defined on connected open subsets of $X_{H}$
such that $x . \xi \neq 0$, where $\xi \in \setR^5 $ lies on the
null cone ${\cal C} = \{ \xi \in \setR^5;\;\; \xi^2=0\}$. They are
homogeneous with degree $-\frac{3}{2}\mp i\nu$ on ${\cal C}$ and
thus are entirely determined by specifying their values on a well
chosen curve (the orbital basis ) $\gamma$ of ${\cal C}$. They can
be written \cite{gata} as a product of a generalized polarization
vector ${\cal E}_{\alpha}(x,\xi,Z_2)$ with the so-called
\cite{brmo}(scalar) dS waves $\left(Hx\cdot\xi\right)^{\sigma}$
where $\sigma=-\frac{3}{2}- i\nu\in \setC $. As such, the dS waves
are multivalued and it will be explained later how suitable
analyticity criteria yield univalued defined waves. The solutions
to Eq. $(\ref{eqn:eqa})$ read
\begin{equation}\label{eq:vector}
K_{\alpha}(x)=\left(\frac{\sigma}{\sigma+1}\right){\cal
E}_{\alpha}(x,\xi,Z_2)(Hx.\xi)^{\sigma},
\quad\mbox{with}\quad\sigma=-\frac{3}{2}- i\nu,
\end{equation}
where $Z_2$ is another constant vector.  Note that contrary to the
Minkowskian case, the polarization tensor is a function of
space-time. The simplest form of ${\cal E}_{\alpha}(x,\xi,Z_2)$
compatible with the Minkowski polarization vector in the flat
limit (see \cite {gata}) is obtained through the choice $\xi\cdot
Z_{2}=0$ and reads
\begin{equation}
{\cal E}(x,\xi,Z_2)=
\left(\bar{Z_{2}}^{\lambda}-\frac{Z_{2}^{\lambda}\cdot x}{x\cdot
\xi}\,\bar{\xi}\right)\quad\mbox{with}\quad{\cal
E}^{\lambda}(x,\xi,Z_2)\cdot \bar\xi=Z_{2}^{\lambda}\cdot\xi =0
\,. \label{eq:vect}
\end{equation}
It is easy to see (flat limit) that the three  Minkowski
polarization four-vectors $\epsilon_{\mu}^{\lambda}$ with
$\mu=0,1,2,3$ are linked to $Z^{\lambda}_2$ by:
\begin{equation}\label{eq:vectlim}
\lim_{H\rightarrow 0}{\cal
E}_{\alpha}^{\lambda}(x,\xi,Z_2)=Z_{2\mu}^{\lambda}-\frac{Z_{24}^{\lambda}}{\xi_{4}}\xi_{\mu}\equiv\epsilon_{\mu}^{\lambda}.
\end{equation}
We demand that the Minkowski polarization  vectors satisfy the
usual relations
\begin{equation}
\epsilon^{\lambda}\cdot
k=0,\quad\epsilon^{\lambda}\cdot\epsilon^{\lambda'}=\eta^{\lambda\lambda'}
,\quad\sum_{\lambda=1}^{3}\epsilon_{\mu}^{\lambda}(k)\,\epsilon_{\nu}^{\lambda}(k)=-\left(\eta_{\mu\nu}-\frac{
k_{\mu}k_{\nu}}{m^{2}}\right)\equiv \Pi_{\mu\nu}(k),
\end{equation}
which is achieved if the $Z_2^{\lambda}$'s are such that
\begin{equation}\label{eq:pola1}
Z_{2}^{\lambda}\cdot\xi=0,\quad Z_{2}^{\lambda}\cdot
Z_{2}^{\lambda'}=\eta^{\lambda\lambda'},\quad\sum_{\lambda=1}^{3}Z^{\lambda}_{2\alpha}Z^{\lambda}_{2\beta}
=-\eta_{\alpha\beta}\qquad\mbox{and}\qquad
\sum_{\lambda=1}^{3}Z^{\lambda}_{24}Z^{\lambda}_{2\mu}=0\;\forall\;\mu\,.
\end{equation}
These conditions are easily derived by working with a well adapted
(to the flat limit) orbital basis. This basis, characterized by
the values $\pm 1$ of the component $\xi_{4}$ will be discussed
later on. A remarkable feature connected with the use of ambient
space notations is that with Eq. (\ref{eq:pola1}) one shows that
the properties of the dS polarization vector are very similar to
the Minkowskian case:
\begin{eqnarray}\label{eq:pola2}
&&\sum_{\lambda=1}^{3}{\cal E
}^{\lambda}_{\alpha}(x,\xi,Z_2)\,{\cal E
}^{\lambda}_{\beta}(x,\xi,Z_2)=-\left(\theta_{\alpha\beta}-\frac{\bar\xi_{\alpha}
\bar\xi_{\beta}}{(Hx\cdot\xi)^2}\right)\equiv \Pi_{\alpha\beta}(x,\xi)\,,\nonumber\\
&&{\cal E }^{\lambda}(x,\xi,Z_2)\cdot{\cal E
}^{\lambda'}(x,\xi,Z_2)=Z^{\lambda}_{2}\cdot{\cal E
}^{\lambda'}(x,\xi,Z_2)={\cal E }^{\lambda}(x,\xi,Z_2)\cdot{\cal E
}^{\lambda'}(x',\xi,Z_2)=\eta^{\lambda\lambda'} \,.
\end{eqnarray}
It follows from Eq. (\ref{eq:wave2}) that the two spin-$2$
families of solutions to Eq. ($\ref{eq:wave1}$) read
$$\K(x)=\theta\phi+{\cal S}\bar Z_1K+D_2K_g \equiv {\cal
D}(x,\partial,Z_1,Z_2)(Hx\cdot\xi)^{-\frac{3}{2} \mp i\nu}$$ where
the operator ${\cal D}(x,\partial,Z_1,Z_2)$ is given by
\begin{equation}
\left(\frac{\sigma}{\sigma+1}\right)\left(-\frac{2}{3}\theta
Z_1\cdot+{\cal S} \bar Z_1+\frac{1}{\langle Q_{0}^{(1)}
\rangle}D_2[\,Z_1\cdot\bar\partial-H^2xZ_1\cdot+3H^2 x\cdot Z_1
-\frac{1}{3}H^2 D_1 Z_1\cdot\, ]\right) \,{\cal E}(x,\xi,Z_2).
\label{eq:pro}
\end{equation}
These spin-2 solutions can be brought into the form
\begin{equation}\label{eq:sol}
\K_{\alpha\beta}(x)=a_{\nu}\,{\cal
E}_{\alpha\beta}(x,\xi,Z_1,Z_2)(Hx\cdot\xi)^{\sigma}
\qquad\mbox{and}\qquad \K_{\alpha\beta}^{*}(x)\,
\quad\mbox{with}\quad
a_{\nu}=c_{\nu}\left(\frac{2(\sigma-1)}{\sigma+1}\right).\end{equation}
where the ${\cal E}_{\alpha\beta}$'s are the generalized
polarization tensor components, $c_{\nu}$ is a normalization
constant and where we have again omitted the superscript
$\lambda$. Because of the conditions
$\K_{\alpha\beta}=\K_{\beta\alpha}$, $\partial.\K=0$, and
$x\cdot\K=0$, the $25$ components ${\cal E}_{\alpha \beta}$ reduce
to 5 independent components which correspond precisely to the $2s
+1 =5$ degrees of freedom of a spin-$2$ field.

The arbitrariness due to the introduction of  the constant vectors
$Z_{1},Z_{2}$ in our solution has partly been removed in
($\ref{eq:pola1}$), by comparison with the Minkowski polarization
vector one eventually reaches by going to the flat limit (see
\ref{eq:vectlim}). We now apply the same procedure in order to fix
the value of $Z_1$, that is we investigate the behaviour of
equation (\ref{eq:pro}) in the $H=0$ limit. More precisely,  we
show that ${\cal E}_{\alpha \beta}(x,\xi,Z_1,Z_2)$ contracts
toward the usual Minkowski tensor polarization and takes a simple
form if $Z_1$ is chosen to be equal to $Z_2$ and denoted by $Z$ in
the following. It is a matter of simple calculation to get the de
Sitter polarization tensor starting with Formula (\ref{eq:pro} ):
\begin{equation}
{\cal E}_{\alpha\beta}(x,\xi,Z)\equiv {\cal
E}_{\alpha\beta}^{\lambda\lambda'}(x,\xi)=\frac{1}{2}
 \left[ {\cal S}\,{\cal
E}_{\alpha}^{\lambda}(x,\xi)\; {\cal
E}_{\beta}^{\lambda'}(x,\xi)-\frac{2}{3}
\left(\theta_{\alpha\beta}
-\frac{\bar{\xi}_{\alpha}\bar{\xi}_{\beta}}{\left(Hx\cdot\xi\right)^{2}}\right)
{\cal E}^{\lambda}(x,\xi)\cdot{\cal E}^{\lambda'}(x,\xi)\right]\,,
\end{equation}
where ${\cal E }^{\lambda}(x,\xi) ={\cal E
}^{\lambda}(x,\xi,Z_2)$. In view of (\ref{eq:pola2}) one obtains
\begin{equation}\label{eq:tensor1}
{\cal E}_{\alpha\beta}^{\lambda\lambda'}(x,\xi)= \frac{1}{2}\left[
{\cal S}\,{\cal E}_{\alpha}^{\lambda}(x,\xi)\; {\cal
E}_{\beta}^{\lambda'}(x,\xi)+\frac{2}{3}\, \eta^{\lambda\lambda'}
\sum_{\rho}{\cal E }^{\rho}_{\alpha}(x,\xi){\cal E
}^{\rho}_{\beta}(x,\xi) \right].
\end{equation}
It is easy to check that the tensor polarization
(\ref{eq:tensor1}) satisfies the properties
$\eta^{\alpha\beta}{\cal E}_{\alpha\beta}(x,\xi,Z)=0$
(tracelessness), $\bar\xi\cdot {\cal E}_{\alpha\beta}(x,\xi,Z)=0$
and the  relation
\begin{equation}
{\cal E}^{\lambda\lambda'}(x,\xi) \cdot \cdot \,{\cal
E}^{\lambda''\lambda'''}(x,\xi) ={\cal
E}^{\lambda\lambda'}(x',\xi) \cdot \cdot \, {\cal
E}^{\lambda''\lambda'''}(x,\xi)=\left[\eta^{\lambda\lambda''}\eta^{\lambda'\lambda'''}
+\eta^{\lambda\lambda'}\eta^{\lambda''\lambda'''}\right]\,.
\end{equation}
The dS tensor waves ${\K}_{\alpha\beta}(x)$ are homogeneous with
degree $\sigma$ on the null cone ${\cal C}$ and on the dS
submanifold $X_{H}$ characterized by $x\cdot x=-H^{-2}$ with $H$
being constant. This is due to:
\begin{equation}
{\cal E}^{\lambda}(x,a\xi)={\cal
E}^{\lambda}(x,\xi)\quad\mbox{and}\quad {\cal
E}^{\lambda}(ax,\xi)={\cal E}^{\lambda}(x,\xi), \nonumber
\end{equation}
which is obvious from the definition of ${\cal
E}^{\lambda}(x,\xi)$
\begin{equation}
{\cal
E}^{\lambda}(x,\xi)=\left(\bar{Z}^{\lambda}-\frac{Z^{\lambda}\cdot
x}{\xi\cdot x
}\,\bar{\xi}\right)=\left({Z}^{\lambda}-\frac{Z^{\lambda}\cdot
x}{\xi\cdot x}\,{\xi}\right).
\end{equation}
Note that as a function of $\setR^{5}$, the wave
${\K}_{\alpha\beta}(x)$ is homogeneous with degree zero
($H(x)=-1/\sqrt{- x\cdot x }$).
\subsection{Flat limit and analytic tensor wave}
It order to compute the flat limit of the polarization tensor, it
is useful to precise the notion of orbital basis $\gamma$ for the
future null cone ${\cal C}^+=\{ \xi \in {\cal C};\; \xi^0>0\},$
\cite{brmo}. Let us choose a unit vector $e$ in $\setR ^{5}$ and
let $H_{e}$ be its stabilizer subgroup in $S0_{0}(1,4)$. Then two
types of orbits are interesting in the present context :
\begin{itemize}
\item[(i)] the spherical type $\gamma_{0}$ corresponds to $e\in V^{+}
\equiv \{ x\in \setR^5;\;\; x^0 >\sqrt{\parallel \vec
x\parallel^2+(x^4)^2} \}$ , and is an orbit of $H_{e}\approx
SO(4)$.
$$\gamma_{0}=\{\xi\;;\;e\cdot\xi=a>0\}\cap  {\cal C}^{+}\;.$$
\item[(ii)] the hyperbolic type  $\gamma_{4}$ corresponds to $e^{2}=-1$.
It is divided into two hyperboloid sheets, both being orbits of
$H_{e}\approx SO_{0}(1,3)$.
\end{itemize}
The most suitable parametrization when one has in view the link
with massive Poincar\'e UIR's is to work with the orbital basis of
the second type
$$\gamma_{4}=\{\xi\;\in C^{+},\xi^{(4)}=1 \}\cup\{\xi\;\in
C^{+},\xi^{(4)}=-1  \},$$ with the null vector $\xi$ given in
terms of the four-momentum $(k^0, \vec{k})$ of a Minkowskian
particle of mass $m$
\begin{equation}
\xi_{\pm}=\left(\frac{k^0}{mc}=\sqrt{\frac{\vec
k^2}{m^2c^2}+1},\frac{\vec k}{mc},\pm 1\right).
\end{equation}
An appropriate choice of global coordinates is given by
\begin{equation}
\left \{\begin{array}{rcllr}
x^0&=&  {\displaystyle H^{-1} \sinh (H X^0) }    \,,\\
\vec x &=&( H\parallel \vec X\parallel)^{-1}\vec X \cosh( HX^0)
\sin
(H\parallel \vec X\parallel) \,,\\
x^4&=&  H^{-1}\cosh( HX^0) \cos (H\parallel \vec X\parallel)\,.
\end{array}
\right.
\end{equation}
where the dS point is expressed in terms of the Minkowskian
variables $X=(X_0=ct, \vec X)$ measured in units of the dS radius
$H^{-1}$.

The Minkowskian limit of the dS  waves at point $x$ can be written
as \cite{brgamo}
\begin{eqnarray}
&&\lim_{H \rightarrow 0}(Hx\cdot\xi_{-})^{\sigma}
=\exp[-ik\cdot X]\; \;\mbox{(positive energy)},\nonumber\\
&&\lim_{H \rightarrow 0}e^{-i\pi \sigma}(Hx\cdot\xi_{+})^{\sigma}
= \exp[ik\cdot X] \;\;\mbox{(negative energy)} \label{eqn:cont}.
\end{eqnarray}
Since the contraction is done with respect to the Lorentz subgroup
$SO_{0}(1,3)$ ($\gamma_{4}$ is invariant under $SO_{0}(1,3)$) the
equations $(\ref{eqn:cont})$ indicate that the orbital basis
$\gamma_{4}$ can contract toward the sum of two solutions with
opposite energies (see \cite{mini}).

The  polarization tensor limit is easily obtained with the help of
$$\lim_{H \rightarrow 0} H^{2}\sigma^{2}=-m^{2},\quad
\lim_{H \rightarrow 0}{\cal
E}_{\alpha}^{\lambda}(x,\xi)=\epsilon_{\mu}^{\lambda}(k)\,,\quad
\lim_{H \rightarrow
0}\theta_{\alpha\beta}=\eta_{\mu\nu}\,,\quad\lim_{H \rightarrow
0}\bar{\xi}_{\alpha}=\frac{k_{\mu}}{m}\;\;\forall\;\;
\xi\in\gamma_{4}.$$ Finally one recovers the Minkowskian massive
spin-2 polarization tensor \cite{pauli}:
$$\lim_{H \rightarrow 0}{\cal E}_{\alpha\beta}^{\lambda\lambda'}(x,\xi)
=\epsilon^{\lambda\lambda'}_{\mu\nu}(k)=\frac{1}{2}{\cal S}\;
\epsilon_{\mu}^{\lambda}(k)\;\epsilon_{\nu}^{\lambda'}(k)
+\frac{1}{3}\eta^{\lambda\lambda'}\sum_{\lambda}\epsilon_{\mu}^{\lambda}(k)\;\epsilon_{\nu}^{\lambda}(k)\,,
$$
which satisfies
$\eta^{\mu\nu}\epsilon^{\lambda\lambda'}_{\mu\nu}(k)=k^{\mu}\epsilon^{\lambda\lambda'}_{\mu\nu}(k)=0$
and
\begin{equation}
\sum_{\lambda\lambda'}\epsilon^{\lambda\lambda'}_{\mu\nu}(k)\epsilon^{\lambda\lambda'}_{\rho\pi}(k)=\frac{1}{2}
\big{[}\Pi_{\mu\rho}(k)\Pi_{\nu\pi}(k)+
\Pi_{\nu\rho}(k)\Pi_{\mu\pi}(k)\big{]}-\frac{1}{3}
\big{[}\Pi_{\mu\nu}(k)\Pi_{\rho\pi}(k)\big{]}\,.
\end{equation}
Hence, we have shown that in the limit $H =0$,
$(Hx\cdot\xi)^{\sigma}$ and ${\cal E}_{\alpha\beta}(x,\xi,Z)$
behave like the plane wave $e^{ik\cdot X}$ and the polarization
tensor in Minkowski space-time respectively.

Although the ``massive'' field equation solutions
$\K_{\alpha\beta}(x)$ and $\K_{\alpha\beta}^{*}(x)$ are complex
conjugated, they  cannot be associated with the positive and
negative energies respectively as in  the Minkowskian situation.
Actually, despite the fact that the solutions are globally defined
(in a distributional sense) in dS space, the concept of energy is
not (absence of global timelike killing vector field). As a
result, concepts like ``particle'' and ``antiparticle'' are rather
unclear and the differences between these two solutions is not
really explained or understood. In terms of group representation
these two solutions are equivalent, because the two
representations $U^{2,\nu}$ and $U^{2,-\nu}$ are.
 Note that the minimally coupled scalar
field requires both sets of solutions in order to achieve a
covariant quantization \cite{gareta1}. This will certainly also be
the case for the spin-2 massless field in dS space since it is
constructed from a minimally coupled scalar field as it will be
shown in \cite{gareta2}.

In the present case, the ``massive'' free field covariant
quantization can be constructed from the positive norm states
alone since $\K_{\alpha\beta}(x)$ is closed under the group
action:
\begin{equation}\label{eq:graction}
\left(U(g) \,\K\right)_{\alpha\beta}(x) = g_{\alpha}^{\gamma}
g_{\beta}^{\delta}\K_{\gamma \delta}(g^{-1} x)=g_{\alpha}^{\gamma}
g_{\beta}^{\delta}\,\,a_{\nu}\,{\cal
E}_{\gamma\delta}(g^{-1}x,\xi,Z)(H g^{-1}x\cdot
\xi)^{\sigma}=a_{\nu}\,{\cal E}_{\alpha\beta}(x,g\xi,gZ)(Hx\cdot
g\xi)^{\sigma}
\end{equation}
This is easily proved since the vector polarization satisfies
\begin{equation}
{\cal E}_{\alpha}(g^{-1}x,\xi,Z)=
\left(Z_{\alpha}-\frac{g^{-1}x\cdot Z}{g^{-1}x\cdot
\xi}\,{\xi_{\alpha}}\right)= \left(Z_{\alpha}-\frac{x\cdot
gZ}{x\cdot
g\xi}\,{\xi_{\alpha}}\right)=(g^{-1})_{\alpha}^{\delta}{\cal
E}_{\delta}(x,g\xi,gZ)\,.
\end{equation}
The dS waves solutions, as functions on de Sitter space,
 are only locally defined since they
are singular on specific lower dimensional subsets of $X_{H}$, for
instance on spatial boundary defined by $x^0 = \pm x^4
\Leftrightarrow x_1^2 + x_2^2 +x_3^2  = H^{-2} $, and multivalued
on dS space-time. In order to get a global definition, they have
to be viewed as distributions \cite{gesh} which are boundary
values of analytic continuations of the solutions to tubular
domains in the complexified de Sitter space $X_H^{(c)}$. The
latter are  defined as follows:
\begin{eqnarray}
X_H^{(c)}&=&\{z=x+iy\in  \setC^5;\;\;\eta_{\alpha \beta} z^\alpha
z^\beta=(z^0)^2-\vec z.\vec
z-(z^4)^2=-H^{-2}\}\nonumber\\
&=&\{ (x,y)\in\setR^5\times \setR^5;\;\; x^2-y^2=-H^{-2},\; x\cdot
y=0\}\,.\nonumber
\end{eqnarray}

For an univalued determination, we must introduce the forward and
backward tubes of $X_H^{(c)}$. First of all, let
$T^\pm=\setR^5-iV^\pm$ be the forward and backward tubes in
$\setC^5$. The domain $V^+$(resp. $V^-)$ stems from the causal
structure on $X_H$:
\begin{equation}
V^\pm=\{ x\in \setR^5;\;\; x^0\stackrel{>}{<}\sqrt{\parallel \vec
x\parallel^2+(x^4)^2} \}.
\label{cone}
\end{equation}
We then introduce their respective intersections with $X_H^{(c)}$,
\begin{equation}
{\cal T}^\pm=T^\pm\cap X_H^{(c)},
\end{equation}
which are the tubes of $X_H^{(c)}$. Finally we define the
``tuboid'' above $X_H^{(c)}\times X_H^{(c)}$ by
\begin{equation}
{\cal T}_{12}=\{ (z,z');\;\; z\in {\cal T}^+,z'\in {\cal T}^- \}.
\end{equation}
 Details are given in \cite{brmo}. When $z$
varies in ${\cal T}^+$ (or ${\cal T}^-$) and $\xi$ lies in the
positive cone ${\cal C}^+$ the  wave solutions are globally
defined because the imaginary part of $(z.\xi)$ has a fixed sign
and $z.\xi\neq 0$.

We define the  de Sitter tensor wave ${\K}_{\alpha\beta}(x)$ as
the boundary value of the analytic continuation to the future tube
of Eq. (\ref{eq:sol}). Hence, for $z \in {\cal T}^+ $ and $\xi \in
{\cal C}^+$ one gets the two solutions
\begin{equation}\label{eq:dswave}
{\K}_{\alpha\beta}(z)= a_{\nu}\,{\cal
E}_{\alpha\beta}^{\lambda\lambda'}(z,\xi) \left( Hz\cdot
\xi\right)^{\sigma},\quad\mbox{and}\quad
{\K}_{\alpha\beta}^{*}(z^{*})=a_{\nu}^{*}\,{\cal
E}_{\alpha\beta}^{*\lambda\lambda'}(z^{*},\xi) \left( Hz\cdot
\xi\right)^{\sigma^{*}}.
\end{equation}
\setcounter{equation}{0}
\section{Two-point function and Quantum field }
\subsection{The two-point function}

As explained in \cite{brmo}, the dS axiomatic field theory is
based on the Wightman two-point double tensor-valued function
\begin{equation}
{\cal W}_{\alpha\beta\alpha'\beta'}(x,x')\qquad
\alpha',\beta'=0,1,..,4. \label{eq:def}
\end{equation}
Indeed, this kernel entirely encodes the theory of the generalized
free fields on dS space-time $X_H$, at least for the massive case.
For this, it has to satisfy the following requirements:
\begin{enumerate}
\item[a)] {\bf Positiveness}

for any test function $f_{\alpha \beta}\in {\cal
D}(X_H)$, we have
\begin{equation}
\int _{X_H \times X_H}f^{*\alpha\beta}(x) {\cal
W}_{\alpha\beta\alpha'\beta'}(x,x')
f^{\alpha'\beta'}(x')d\sigma(x)d\sigma(x')\geq 0,
\end{equation}
where $d\sigma (x)$ denotes the dS-invariant measure on $X_H$
\cite{brmo}. ${\cal D}(X_H)$ is the space of functions $C^\infty$
with compact support in $X_H$.

\item[b)] {\bf Locality}
for every space-like separated pair $(x,x')$, {\it i.e.} $x\cdot
x'>-H^{-2}$,
\begin{equation}
{\cal W}_{\alpha\beta \alpha'\beta'}(x,x')={\cal
W}_{\alpha'\beta'\alpha\beta }(x',x) .
\end{equation}
\item[c)] {\bf Covariance}
\begin{equation}
(g^{-1})^{\gamma}_{\alpha}(g^{-1})^{\delta}_{\beta} {\cal
W}_{\gamma\delta \gamma'\delta'} (g x,g x')g^{\gamma'}_{\alpha'}
g^{\delta'}_{\beta'}= {\cal W}_{\alpha\beta \alpha'\beta'}(x,x'),
\end{equation}
for all $g\in$ SO$_0(1,4)$.
\item[d)] {\bf Index symmetrizer}
\begin{equation}
{\cal W}_{\alpha\beta \alpha'\beta'}(x,x')={\cal W}_{\alpha\beta
\beta'\alpha '}(x,x')={\cal W}_{\beta\alpha\alpha'\beta'}(x,x').
\end{equation}
\item[e)] {\bf Transversality}
\begin{equation}
x^\alpha {\cal W}_{\alpha\beta \alpha'\beta'}(x,x')=0=x'^{\alpha'}
{\cal W}_{\alpha\beta\alpha'\beta'}(x,x') .
\end{equation}
\item[f)] {\bf Divergencelessness}
\begin{equation} \partial_x^\alpha {\cal
W}_{\alpha\beta
\alpha'\beta'}(x,x')=0=\partial_{x'}^{\alpha'}
{\cal W}_{\alpha\beta \alpha'\beta'}(x,x').
\end{equation}
\item[g)] {\bf Normal analyticity}
${\cal W}_{\alpha\beta \alpha'\beta'}(x,x')$ is the boundary value
(bv) in the distributional sense of an analytic function
$W_{\alpha\beta \alpha'\beta'}(z,z').$
\end{enumerate}

Concerning the last requirement,
$W_{\alpha\beta\alpha'\beta'}(z,z')$ is actually maximally
analytic, {\it i.e.} can be analytically continued to the ``cut
domain''
$$\Delta=\{(z,z') \in X_H^{(c)} \times X_H^{(c)} \;\; :\;\;(z-z')^2< 0\}.$$
The Wightman  two-point function ${\cal
W}_{\alpha\beta\alpha'\beta' }(x,x')$ is the boundary value of
$W_{\alpha\beta\alpha'\beta'}(z,z')$ from ${\cal T}_{12}$ and the
``permuted Wightman function'' ${\cal W}_{\alpha'\beta'\alpha\beta
}(x',x)$ is the boundary value of
$W_{\alpha\beta\alpha'\beta'}(z,z')$ from the domain
$$ {\cal T}_{21}=\{ (z,z');\;\; z\in {\cal T}^-, z'\in{\cal T}^+\}. $$
Once these properties are satisfied, the reconstruction theorem
\cite{stwi}  allows to recover the corresponding quantum field
theory. Our present task is therefore to find a doubled tensor
valued analytic function of the variable $ (z, z')$  satisfying
the properties a) to g). Following Reference \cite{brmo} (in which
the construction has been done for the scalar case), the analytic
two-point function $W_{\alpha\beta\alpha'\beta'}(z,z') \equiv
W^\nu _{\alpha\beta\alpha'\beta'}(z,z')$ is obtained from the dS
tensor waves  $(\ref{eq:dswave})$. The parameter $\nu$ refers to
the principal series. The two-point function is given in terms of
the following class of integral representations
\begin{equation}
W^\nu _{\alpha\beta \alpha'\beta'}(z,z')= \vert
a_{\nu}\vert^{2}\int_\gamma (Hz\cdot \xi)^{\sigma} (H
z'\cdot\xi)^{\sigma^{*}}\sum_{\lambda\lambda'}{\cal
E}^{\lambda\lambda'}_{\alpha\beta }(z,\xi) \,{\cal
E}^{*\lambda\lambda'}_{\alpha'\beta'}(z^{'*},\xi)\,d\sigma_\gamma(\xi),
\label{eq:int}
\end{equation}
where  $d\sigma_{\gamma}(\xi)$ is the natural ${\cal C}^{+}$
invariant measure on $\gamma$, induced from the $\setR^{5}$
Lebesgue measure \cite{brmo} and the normalization constant
$a_{\nu}$ is fixed by local Hadamard condition. The latter selects
a unique vacuum state for quantum tensor fields which satisfies
the dS field equation. In order to check wether condition a) to g)
are satisfied by Eq. (\ref{eq:int}) let us first  rewrite  the
two-point function in a more explicit way. This will be done by
using the scalar and the vector ``massive'' analytic two-point
functions  $W^{\nu}_{0}(z,z')$, $ W_{1}^{\nu}(z,z')$ (where
$\z=-H^{2}z\cdot z'$). The latter satisfy the complex versions of
the Casimir equations:
\begin{equation}
\left(Q_{1}-\langle Q_{1}^{(1)}\rangle\right) W^{\nu}_{1}(z,z')=0
\label{eq:tp1}\qquad\mbox{and}\qquad\left(Q_{0}-\langle
Q_{0}^{(1)}\rangle\right)W^{\nu}_{0}(z,z')=0\, .
\end{equation}
In appendix C and in Reference \cite{gata} it is shown how $
W^{\nu}_{1}(z,z')$ can be written in terms of the scalar analytic
two-point function
\begin{equation}
W^{\nu}_{1}(z,z') =\frac{\langle Q_{0}\rangle }{\langle
Q_{1}\rangle}\left(-\,\theta_{\alpha}\cdot\theta'_{\alpha'}+
\frac{H^{2}\sigma(\theta\cdot z' )D'_{1}}{\langle Q_{0}\rangle }+
\frac{H^{2}\sigma^{*}(\theta'\cdot z) D_{1}}{\langle Q_{0}\rangle
}+ \frac{ H^2\z D_{1}D'_{1}}{\langle Q_{0}\rangle}\right)
W^{\nu}_{0}(z,z').\label{eq:vec}
\end{equation}
The Wightman scalar two-point function ${\cal W}^{\nu}_{0}(x,x')$
is given by \cite{brmo}
\begin{equation}\label{eq:sca}
{\cal
W}^{\nu}_{0}(x,x')=\mbox{bv}\;W_{0}(z,z')\qquad\mbox{with}\qquad
W^\nu_{0}(z,z')= c_\nu^{2}\int_\gamma (Hz\cdot\xi)^{\sigma}(H
z'\cdot \xi )^{\sigma^{*}}\,d\sigma_\gamma(\xi)\,.
\end{equation}
The normalization constant $c_{\nu}^{2}$ is determined by imposing
the Hadamard condition on the two-point function. This has been
done in Ref. \cite{brmo} where the scalar two-point function has
been rewritten in terms of the generalized Legendre function  for
well chosen space like separated points $z$ and $z'$. It has been
established that $W_{0}(z,z') = C_\nu P_{\sigma}^{(5)}(-\z)$ with
$C_{\nu}=2\pi^2 e^{-\pi \nu} c_\nu^{2}$ and
\begin{equation}
c_\nu^{2}=\frac{H^2e^{\pi\nu}\Gamma(-\sigma)\Gamma(-\sigma^{*})}{2^5\pi^4
m^{2}}. \label{eq:nor}
\end{equation}

This normalization corresponds to the Euclidean vacuum \cite{brmo}
and $P_{\sigma}^{(5)}(\z)$ is the generalized Legendre function of
the first kind. There are several reasons which explain the
appearance of $W^{\nu}_{0}(z,z')$ and $ W_{1}^{\nu}(z,z')$. First
of all, both correspond to the commonly used two-point functions
(see for instance reference \cite{allen}) as it is checked in
Appendix C. Moreover, since the vector two-point function is
written in terms of the scalar two-point function it exhibits the
two building blocks of the tensor expression which are well known
and simple to manipulate. As a matter of fact, the flat limit is
very easy to compute in this framework.

We have seen that the  spin-2 analytic two-point function
(\ref{eq:int}) is obtained from the tensor waves
(\ref{eq:dswave}). Let us cast the latter into the more suitable
form
\begin{equation}
{\cal K}(z)=\frac{a_{\nu}}{2} \left[ {\cal S}\;{\cal
E}^{\lambda}(z,\xi){\cal E}^{\lambda'}(z,\xi)-\frac{2\sigma
g^{\lambda\lambda'}}{3(\sigma-1)}\left(
\theta-\frac{H^{2}D_{2}D_{1}}{2\sigma^{2}}\right) \right]\left(H
z\cdot \xi \right)^{\sigma}\,,
\end{equation}
by using the property
\begin{equation}
\sum_{\lambda}{\cal E }^{\lambda}(z,\xi)\,{\cal E
}^{\lambda}(z,\xi)\left( H z\cdot\xi
\right)^{\sigma}=-\left(\theta-\frac{\bar\xi
\bar\xi}{(Hz\cdot\xi)^{2}}\right)\left( H z\cdot\xi
\right)^{\sigma}=-
\frac{\sigma}{\sigma-1}\left[\theta-\frac{H^{2}D_{2}D_{1}}{2\sigma^{2}}\right]\left(
H z\cdot\xi \right)^{\sigma}\,.
\end{equation}
We then simply develop the two-point function and obtain :
\begin{eqnarray}\label{eqn:tpfunction1}
W^{\nu}(z,z')&=& \frac{\vert a_{\nu}\vert^{2}}{4}\,
\int_\gamma{\cal S }{\cal S' }\left(\sum_{\lambda}{\cal
E}^{\lambda}(z,\xi) \,{\cal
E}^{*\lambda}(z^{'*},\xi)\right)\left(\sum_{\lambda'}{\cal
E}^{\lambda'}(z,\xi) \,{\cal
E}^{*\lambda'}(z^{'*},\xi)\right)(Hz\cdot\xi)^{\sigma} (H z'\cdot \xi)^{\sigma^{*}}\,d\sigma_\gamma(\xi)\nonumber\\
&-&\frac{4}{3}\frac{\langle Q_{0}\rangle }{\langle Q_{1}\rangle
}\left[ \theta-\frac{H^{2}D_{2}D_{1}}{2\sigma^{2}}\right] \left[
\theta'-\frac{H^{2}D'_{2}D'_{1}}{2\sigma^{*2}}\right]c_\nu^{2}\int_\gamma
(Hz\cdot\xi)^{\sigma} (H z'\cdot \xi
)^{\sigma^{*}}\,d\sigma_\gamma(\xi)\,.
\end{eqnarray}
From  the  property
\begin{equation}\label{eq:prop1}
\sum_{\lambda}{\cal E}^{\lambda}(z){\cal
E}^{*\lambda}(z'^{*})=\left[-
\theta\cdot\theta'+\frac{(\theta\cdot
z')\bar\xi'}{z'\cdot\xi}+\frac{(\theta'\cdot
z)\bar\xi}{z\cdot\xi}+\frac{\z\,\bar\xi\,\bar\xi'}{H^{2}z\cdot\xi
z'\cdot\xi}\right]\,,
\end{equation}
and  the relation $H^{2}\,D_{2}\,K(x)=(\sigma-1)\,{\cal S
}\,\bar\xi \,K (x)/ \left(z\cdot\xi\right)\,, $ it is clear that
the analytic two-point function can be written in  the general
form:
\begin{eqnarray}\label{eqn:tpfunction2}
W^{\nu}(z,z')= M(z,z')\,W_{1}^{\nu}(z,z')
+N(z,z')\,W^{\nu}_{0}(z,z')\,.
\end{eqnarray}
The differential operators $M(z,z')$ and $N(z,z')$ are given by
\begin{eqnarray}\label{eqn:fetg}
M(z,z')&=&\frac{\langle Q_{0}\rangle+4}{\langle Q_{0}\rangle
}\left[-\,{\cal S }{\cal S' } \theta\cdot\theta'+\frac{H^{2}{\cal
S }(\theta\cdot z')D'_{2}}{\sigma^{*}-1}+\frac{H^{2}{\cal S'
}(\theta'\cdot z)D_{2}}{\sigma-1}+\frac{\z H^{2}D_{2}D'_{2}}{
(\sigma-1)(\sigma^{*}-1)}\right]\,,
\nonumber\\
N(z,z')&=&\frac{4}{3}\frac{\langle Q_{0}\rangle }{\langle
Q_{1}\rangle }\left[
\theta-\frac{H^{2}D_{2}D_{1}}{2\sigma^{2}}\right]\left[
\theta'-\frac{H^{2}D'_{2}D'_{1}}{2\sigma^{*2}}\right]\,.
\end{eqnarray}
Eventually, the analytic tensor two-point function is given in
terms of the scalar analytic two-point function by:
$$W^{\nu}_{\alpha\beta
\alpha'\beta'}(z,z')=D(z,z') W_{0}^{\nu}(z,z')\;,$$ with $D(z,z')$
a differential operator discussed in appendix D. The boundary
value of $W^\nu (z,z')$ gives the following integral
representation for the Wightman two-point function:
\begin{equation}
 {\cal W}(x,x')=
\vert
a_{\nu}\vert^{2}\sum_{\lambda\lambda'}\int_{\gamma}d\sigma_{\gamma}(\xi)
{\cal E}^{\lambda\lambda'}(x,\xi) {\cal
E}^{*\lambda\lambda'}(x',\xi) \,\mbox{bv} \left(H
z\cdot\xi\right)^{\sigma} \left(H z'\cdot\xi\right)^{\sigma^*} \,,
\end{equation}
with
\begin{equation}
\mbox{bv} \left(H z\cdot\xi\right)^{\sigma} \left(H
z'\cdot\xi\right)^{\sigma^*}= \vert H x\cdot\xi\vert^{\sigma}
\vert H x'\cdot\xi\vert^{\sigma^*} \big{[}\theta(H x\cdot \xi) +
\theta(-H x\cdot\xi)\,e^{-i\pi \sigma } \big{]} \big{[} \theta(H
x'\cdot \xi) + \theta(-H x'\cdot\xi)\,e^{+i\pi \sigma^{*} }
\big{]}.
\end{equation}
This relation defines the two-point function in terms of global
waves on the real hyperboloid $X_H$.

Let us now check if this kernel fulfills the conditions a) to g)
required in order to get a Wightman  two-point function. We recall
that  the existence of the latter which is requested by dS
axiomatic field theory.
\begin{itemize}
\item The positiveness property follows from the relation
\begin{equation}
\int _{X_H \times X_H}f^{*\alpha\beta}(x) {\cal
W}_{\alpha\beta\alpha'\beta'}(x,x')
f^{\alpha'\beta'}(x')d\sigma(x)d\sigma(x')=\vert
a_{\nu}\vert^{2}\int_{\gamma}d\sigma_{\gamma}(\xi)\sum_{\lambda\lambda'}
g^{*\lambda\lambda'}(\xi)\,g^{\lambda\lambda'}(\xi)\,,
\end{equation}
where
\begin{equation}
g^{\lambda\lambda'}(\xi)=\int_{X_H}
d\sigma(x)f^{\alpha\beta}(x){\cal
E}^{*\lambda\lambda'}_{\alpha\beta}(x,\xi)\big{[} \theta(H x\cdot
\xi) + \theta(-H x\cdot\xi)\,e^{+i\pi \sigma^{*} } \big{]}\vert H
x\cdot\xi\vert^{\sigma^{*}} \,.
\end{equation}
The hermiticity property is obtained,  by considering boundary
values of the following identity
\begin{equation}
W_{\alpha\beta
\alpha'\beta'}(z,z')=W_{\alpha'\beta'\alpha\beta}^*(z^{\prime*},z^*),
\end{equation}
which is easily checked on Eq.
($\ref{eq:int}$).

\item In order to prove the locality condition, we use  the hermiticity
condition and the following relation:
$$ W^{*}_{\alpha'\beta'\alpha\beta
}(z'^*,z^*)=W_{\alpha'\beta' \alpha\beta }(z',z).$$ This  easily
follows from the form of the two-point function for space-like
separated points given in Appendix D :
$$W^{\nu}(z,z')=C_{\nu}D(z,z')P_{\sigma}^{(5)}(-\z)\quad\mbox{with}\quad
D^{*}(z^*,z'^{*})=D(z,z'),$$ and from the  relation \cite{brvia}
$$P_{\sigma}^{(5)}(-\z) =P_{\sigma^{*}}^{(5)}(-\z). $$ One finally gets
$$W_{\alpha\beta\alpha'\beta'
}(z,z')=W_{\alpha'\beta'\alpha\beta}^*(z^{\prime*},z^*)=W_{\alpha'\beta'
\alpha\beta }(z',z).
$$
It should be noticed that the space-like separated pair ($x,x'$)
lies in the same orbit of the complex dS group as the pairs
($z,z'$) and ($z'^{*},z^*$). Therefore the locality condition
${\cal W}_{\alpha\beta \alpha'\beta'}(x,x')={\cal
W}_{\alpha'\beta' \alpha\beta}(x',x)$ holds.

\item The group action on the dS modes (\ref{eq:graction})
and the independence of the integral  ($\ref{eq:int}$) with
respect to the selected orbital basis  entail the covariance
property
\begin{equation}
(g^{-1})^{\gamma}_{\alpha}(g^{-1})^{\delta}_{\beta} {\cal
W}_{\gamma\delta \gamma'\delta'} (g x,g x')g^{\gamma'}_{\alpha'}
g^{\delta'}_{\beta'}= {\cal W}_{\alpha\beta \alpha'\beta'}(x,x') .
\end{equation}
\item The symmetry with respect to the indices $\alpha$, $\beta$
and $\alpha'$, $\beta'$ and the transversality with respect to $x$
and $x'$ are guaranteed by construction. So is the divergencelessness
condition.
\item The analyticity properties of the tensor Wightman two-point
function follow from the expression of the dS tensor waves
($\ref{eq:dswave}$).
\end{itemize}
{\bf Remark}

A massive spin-2 two-point function had already been proposed in
Ref. \cite{gasp}. Although the approach we have used here is very
different (in Ref. \cite{gasp} the coordinates are non global, the
modes have a spin-0 and spin-2 content..) it has been possible to
check that our vector two-point function is in agreement with the
one presented in \cite{gasp}. This is of importance since it
confirms for tensor fields the validity of the integral
representation method (\ref{eq:int}) originated in \cite{brmo} for
the scalar case. However, explicit  comparison for the spin-2 case
would be a rather tedious task given the differences between both
formalisms and the involved expression of the spin-2 two-point
function given in \cite{gasp}. It seems that one can at least say
that the ambient space formalism presents the advantage of
simplicity. This is again verified by performing the flat limit as
it is seen in the next paragraph and this was already the case
when the unitary irreducible representations had to be identified
in section II.

\subsection{The flat limit}
The flat limit is straightforward to compute with  the help of the
orbital basis $\gamma_{4}$. The measure
$d\sigma_{\gamma_{4}}(\xi)$  is chosen to be  $m^2$ times the
natural one induced from the $\setR^{5}$ Lebesgue measure. This
yields $d\sigma_{\gamma_4}(\xi)=d^{3}\vec{k}/k_{0}$ and the
constant $\vert a_{\nu}\vert^{2}$ reads
\begin{equation}
\vert a_{\nu}\vert^{2}=4\frac{\langle Q_0 \rangle+4 }{\langle Q_1
\rangle}\left[ \frac{H^2 e^{\pi\nu}\Gamma(-\sigma)
\Gamma(-\sigma^{*})}{2^5\pi^4m^2}\right]=4\frac{\langle Q_0
\rangle+4 }{\langle Q_1 \rangle}\left[
\frac{H^2\nu^{2}+H^{2}/4}{2^4\pi^3m^2}\right]\,.
\end{equation}

One finds the massive spin-$2$  Minkowski two-point function:
\begin{eqnarray}
\lim_{H\rightarrow 0}\frac{1}{4}{\cal W}^\nu(x,x')=
\frac{1}{{2(2\pi)^3}} \int
\sum_{\lambda\lambda'}\epsilon^{\lambda\lambda'}(k)\epsilon^{\lambda\lambda'}(k)
\exp(-ik(x-x'))d^{3}\vec{k}/k_{0} \,,
\end{eqnarray}
where the factor $1/4$ is due to our definition of the operators
${\cal S}$ and ${\cal S}'$. This limit can also be computed (more
explicitly) starting with Formula (\ref{eqn:tpfunction2}). The
flat limit for the scalar and vector two-point functions have been
computed in \cite{brmo,gata}, one obtains:
\begin{equation}
\lim_{H\rightarrow 0}{\cal W}^{\nu}_{0}(x,x')={\cal
W}^{P}(X,X'),\quad\lim_{H\rightarrow 0}{\cal
W}_{1}(x,x')=-\left[\eta_{\mu\nu}+\frac{1}{m^{2}}
\frac{\partial}{\partial X^{\mu}\partial X^{\nu}}\right] {\cal
W}^{P}(X,X')\equiv{\cal W}_{\mu\nu}^{P}(X,X')\,,
\end{equation}
where ${\cal W}^{P}(X,X')$ and ${\cal W}_{\mu\nu}^{P}(X,X')$ are
the scalar and vector massive  Minkowskian two-point functions
respectively. Under the constraint $H\nu=m$ which implies
\begin{equation}
\lim_{H\rightarrow 0}H^{2}\langle
Q_{s}\rangle=m^{2}\quad\mbox{and}\quad \lim_{H\rightarrow
0}H^{2}\sigma^{2}=-m^{2}\,,
\end{equation}
one finally gets the massive spin-$2$ Minkowski two-point function
(see for instance \cite{higu1})
\begin{equation}
\lim_{H\rightarrow 0}\frac{1}{4}{\cal W}(x,x')= +\frac{1}{3}
\left[ \eta_{\mu\nu}+\frac{1}{m^{2}} \frac{\partial}{\partial
X^{\mu}\partial X^{\nu}}\right]{\cal
W}_{\mu'\nu'}^{P}(X,X')-\frac{1}{2}{\cal S} \left[
\eta_{\mu\nu'}+\frac{1}{m^{2}} \frac{\partial}{\partial
X^{\mu}\partial X^{\nu'}}\right] {\cal W}_{\mu'\nu}^{P}(X,X')\,.
\end{equation}

\subsection{The quantum field}
 The explicit
knowledge of ${\cal W}^{\nu}(x,x')$ allows us to make the QF
formalism work. The tensor fields ${\bf{\cal K}}(x)$ is expected
to be an operator-valued distributions on $X_H$ acting on a
Hilbert space ${\cal H}$. In terms of Hilbert space and field
operator, the properties of the Wightman two-point functions are
equivalent to the following conditions \cite{stwi}:
\begin{enumerate}
\item {\bf Existence of an unitary irreducible
representation of
the dS group}
 $$ U = U^{2,\nu},\;\; (\mbox{and possibly} \;\; V^{2,q}),$$
\item {\bf Existence of at least one ``vacuum state''}
 $\Omega$,
cyclic for the polynomial algebra of field operators
and invariant
under the above representation of the dS group.
\item {\bf Existence of a Hilbert space} ${\cal H}$
with positive definite metric that can be described as the
Hilbertian sum
$$
{\cal H}={\cal H}_0\oplus[\oplus_{n=1}^{\infty}{\cal S}{\cal
H}_1^{\bigotimes n}],
$$
where ${\cal H}_0=\{ \lambda \Omega,\;\; \lambda \in \setC\}$.
\item {\bf Covariance }
of the field operators under the representation $U$,
$$U(g){ \bf {\cal K}}_{\alpha\beta}(x) U(g^{-1})=
g_{\alpha}^{\gamma} g_{\beta}^{\delta}{\cal K}_{\gamma \delta}(g
x).$$

\item{\bf Locality}
for every space-like separated pair $(x,x')$
$$[{\cal K}_{\alpha\beta}(x),{\cal K}_{\alpha'\beta'}(x')]=0. $$

\item {\bf KMS condition or geodesic spectral
condition} \cite{brmo}
which means the vacuum is defined as a physical state
with the
temperature $T=\frac{H}{2\pi}$.

\item {\bf Transversality}
$$ x\cdot {\cal K}(x)=0.$$

\item {\bf Divergencelessness}
$$ \partial\cdot {\cal K}(x)=0.$$

\item {\bf Index symmetrizer}
$$  {\cal K}_{\alpha \beta}={\cal K}_{\beta \alpha}. $$
\end{enumerate}
Given the two-point function, one can realize the Hilbert space as
functions on $X_{H}$ as follows. For any test function $f_{\alpha
\beta}\in {\cal D}(X_H)$, we define the vector valued distribution
taking values in the space generated by the modes
$\K_{\alpha\beta}(x,\xi)\equiv \mbox{bv}\,\K_{\alpha\beta}(z,\xi)$
by :
\begin{equation}
x\rightarrow p_{\alpha\beta}(f)(x)=\int_{X_{H}} {\cal
W}_{\alpha\beta\alpha'\beta'}(x,x')
f^{\alpha'\beta'}(x')d\sigma(x')=\sum_{\lambda\lambda'}\int_{\gamma}d\sigma_{\gamma}(\xi)
{\K}^{\lambda\lambda'}_{\xi}(f)\,{\K}_{\alpha\beta}^{\lambda\lambda'}(x,\xi)\,,
\end{equation}
where $\K^{\lambda\lambda'}_{\xi}(f)$ is the smeared form of the
modes:
\begin{equation}
{\K}^{\lambda\lambda'}_{\xi}(f)=\int_{X_{H}}{\K}_{\alpha\beta}^{*\lambda\lambda'}(x,\xi)f^{\alpha\beta}(x)d\sigma(x)\,.
\end{equation}
The space generated by the $p(f)$'s is equipped with the positive
invariant inner product
\begin{equation}
\langle p(f),p(g)\rangle=\int_{X_{H}\times X_{H}}
f^{*\alpha\beta}(x){\cal W}_{\alpha\beta\alpha'\beta'}(x,x')
g^{\alpha'\beta'}(x')d\sigma(x')d\sigma(x)\,.
\end{equation}
As usual, the field is defined by the operator valued distribution
\begin{equation}
{\cal K}(f)=a\left(p(f)\right)+a^{\dagger}\left(p(f)\right)\,,
\end{equation}
where the operators $a(\K^{\lambda\lambda'}(\xi)) \equiv
a^{\lambda\lambda'}(\xi)$ and
$a^{\dagger}(\K^{\lambda\lambda'}(\xi)) \equiv
a^{\dagger\lambda\lambda'}(\xi)$ are respectively antilinear and
linear in their arguments. One gets:
\begin{equation}
{\cal
K}(f)=\sum_{\lambda\lambda'}\int_{\gamma}d\sigma_{\gamma}(\xi)
\left[\K^{*\lambda\lambda'}_{\xi}(f)\,a^{\lambda\lambda'}(\xi)
+\K^{\lambda\lambda'}_{\xi}(f)\,a^{\dagger\lambda\lambda'}(\xi)\right]\,.
\end{equation}
The unsmeared operator reads
\begin{equation}\label{eq:field}
{\cal K}_{\alpha\beta}(x)= \sum_{\lambda \lambda'}\int_{\gamma}
d\sigma_{\gamma}(\xi) \left[ \;{\cal
\K}_{\alpha\beta}^{\lambda\lambda'}(x,\xi)\,
a^{\lambda\lambda'}(\xi) +{\cal
\K}_{\alpha\beta}^{*\lambda\lambda'} (x,\xi)\,
a^{\dagger\lambda\lambda'}(\xi)\right]\,,
\end{equation}
where $a^{\lambda\lambda'}(\xi)$ satisfies the canonical
commutation relations (ccr) and is defined by
$$a^{\lambda\lambda'}(\xi)|\Omega>=0.$$
The measure satisfies
$d\sigma_{\gamma}(l\xi)=l^{3}d\sigma_{\gamma}(\xi)$ and
${\K}^{\lambda\lambda'}_{\alpha\beta}(x,l
\xi)=l^{\sigma}{\K}^{\lambda\lambda'}_{\alpha\beta}(x,\xi)$ yields
the homogeneity condition
$$ \;a^{\lambda\lambda'}(l \xi)\equiv a(\K^{\lambda\lambda'}(l\xi))=a(l^{\sigma}\K^{\lambda\lambda'}(\xi))
=l ^{\sigma^{*}}a^{\lambda\lambda'}(\xi).$$ The integral
representation ($\ref{eq:field}$) is independent of the orbital
basis $\gamma$ as explained in \cite{brmo}.  For the hyperbolic
type submanifold $\gamma_{4}$ the measure is
$d\sigma_{\gamma_{4}}(\xi)=d^{3}\vec{\xi}/\xi_{0}$ and the ccr are
represented by

\begin{equation}
[a^{\lambda\lambda'}(\xi),a^{\dagger\lambda''\lambda'''}(\xi')]=
\left[\eta^{\lambda\lambda''}\eta^{\lambda'\lambda'''}
+\eta^{\lambda\lambda'}\eta^{\lambda''\lambda'''}\right]\xi^0\delta^3(\vec\xi-\vec\xi').
\end{equation}

The field commutation relations are
\begin{equation}
\left[{\cal K}_{\alpha\beta}(x),{\cal
K}_{\alpha'\beta'}(x')\right]=2i \mbox{Im} \langle
p_{\alpha\beta}(x),p_{\alpha'\beta'}(x')\rangle=2i \mbox{Im}{\cal
W}_{\alpha\beta\alpha'\beta'}(x,x') \,.
\end{equation}

\section{Conclusion}

In this paper we have considered the ``massive'' spin-$2$ tensor
field that is associated to the principal series of the dS group
SO$_{0}(1,4)$ with $<Q_\nu>=\nu^2-\frac{15}{4},\;\;\nu \geq 0 $
and corresponding to the nonzero ``mass''
$m_p^2=H^2(\nu^2+\frac{9}{4})$. In our view, the use of the
``mass" concept is more forced by tradition than relevant to our
analysis. The use of ambient space formalism endowed the de Sitter
physics with a Minkowskian-like appearance. The main differences
hold in  the space time dependence of the de Sitter polarization
tensor. This formalism yield simple expressions and make de Sitter
QFT look almost like standard QFT in flat space time.

The group theoretical point of view allows a systematic and
complete study of the spin-2 field theory and legitimates the
restriction of ``massive" fields to those which carry principal
series representations. Indeed, in the case of the complementary
series ($<Q_\mu>=\mu-4,\;\;0<\mu <\frac{1}{4}$), although the
associated ``mass" $m_c^2=H^2(\mu+2),\;\;0<\mu <\frac{1}{4}$ is
strictly positive, the physical meaning of their carrier fields
remains unclear since the $H=0$ limits of these representations in
the complementary series do not correspond to any physical
representation of the Poincar\'e group.

 Since  $m_p^2$ and
$m_c^2$ are strictly  non zero, ``massless" spin-2 fields must
belong to the discrete series among which only $\Pi^\pm_{s,s}$
have a physically meaningful Poincar\'e limit. Now since the
associated ``mass" is $m_d^2=H^2\{6-2(s^{2}-1)\},\;\;s \geq 2$,
and is expected to be real, the only possible value of $s$ is $2$
with $m_d^2=0$. Hence $\Pi^\pm_{2,2}$ correspond precisely to
``massless" tensor fields (linear quantum gravity in dS space) in
perfect agreement with the fact that on  one hand these
representations have non ambiguous extensions to the conformal
group $SO(4,2)$ and on the other hand, the latter are precisely
the unique extensions of the massless Poincar\'e group
representations with helicity $\pm 2$. In this case $\nu $ should
be replaced by $\pm \frac{3i}{2}$ in the formulas of the present
paper \cite{ta}. The projection operator ${\cal D}$ (Eq.
$(\ref{eq:pro}$) on the classical level) and the normalization
constant $c_\nu^{2}$ (Eq. ($\ref{eq:nor}$) on the quantum level)
then become singular. This singularity is actually due to the
divergencelessness condition needed to associate the tensor field
with a specific UIR of the dS group. To solve this problem, the
divergencelessness condition must be dropped. Then the field
equation becomes gauge invariant, {\it i.e.} ${\cal K}^{gt}={\cal
K}+D_2\Lambda_g$ is a solution of the field equation for any
vector field $\Lambda_g$ as far as ${\cal K}$ is. As a result, the
general solutions transform under  indecomposable representations
of the dS group. By fixing the gauge, the field can  eventually be
quantized.

A second type of singularity appears. It is due to the zero mode
problem of the Laplace-Beltrami operator on dS space inherited
from  the minimally coupled scalar field \cite{gareta1}.
Accordingly, we feel that a Krein space quantization along the
lines presented in
  \cite{gareta1} can be successfully
carried out in the spin-2 massless case in dS space. This
situation will be considered in a forthcoming paper
\cite{gareta2}.

\vspace{0.5cm}

\noindent {\bf{Acknowledgements}}: We are grateful to J. Renaud
and S. Rouhani for useful discussions.

\newpage
\setcounter{equation}{0}
\begin{appendix}
\section{Classification of the unitary irreducible representations of
the de Sitter group SO$_{0}(1,4)$.}
 Unitary irreducible representations (UIR) of
SO$_{0}(1,4)$ are characterized by the eigenvalues of the two
Casimir operators $Q^{(1)}$ and $Q^{(2)}$ introduced in Section
II. In fact the UIR's may be labelled by a pair of parameters
$\Delta =(p,q)$ with  $2p \in {\setN}$ and $q \in {\setC}$, in
terms of which the eigenvalues of $Q^{(1)}$ and $Q^{(2)}$ are
expressed as follows \cite{bagamota,dix,tak}:
$$
{Q^{(1)}}=[-p(p+1)-(q+1)(q-2)]{\1}, \quad
{Q^{(2)}}=[-p(p+1)q(q-1)]{\1}.
$$
According to the possible values for $p$ and $q$, three series of
inequivalent representations may be distinguished: the principal,
complementary and discrete series. We write $s$ when $p$ or $q$
have spin meaning.
\begin{enumerate}
\item Principal series representations $U_{s,\nu}$, also called
``massive'' representations:\quad $\Delta=\left(s,{1\over
2}+i\nu\right)$ with
\begin{eqnarray}
&&s=0,1,2,\dots \quad {\rm and} \quad  \nu\geq 0 \quad {\rm
or},\nonumber\\
&&s={1\over 2},{3\over 2},\dots \quad \ \, {\rm and} \quad  \nu
>0.\nonumber
\end{eqnarray}
The operators $Q^{(1)}$ and $Q^{(2)}$ take respectively the
following forms:
$${Q_1} =\Bigl[ \bigl( {9\over 4}+{\nu ^2} \bigr)-s(s+1) \Bigr]\,
{\1},\qquad {Q_2} =\Bigl[ \bigl( {1\over 4}+{\nu ^2} \bigl)s(s+1)
\Bigr]\, {\1}.$$
They are called the massive representations of
the dS group because they contract toward the massive spin $s$
representations of the Poincar\'e group.
\item  Complementary series representations
 $V_{s,\nu}$:\quad $\Delta=(s,{1\over 2}+\nu)$
with
\begin{eqnarray}
&&s=0 \quad {\rm and} \quad  \nu\in \setR\ ,\ 0<|\nu|<{3\over 2}
\quad {\rm
or},\nonumber\\
&&s=1,2,3,\dots \quad {\rm and} \quad \nu\in{\setR}\ ,\
0<|\nu|<{1\over 2}.\nonumber
\end{eqnarray}
The operators $Q^{(1)}$ and $Q^{(2)}$ take forms:
$$
{Q_1} =\Bigl[ \bigl( {9\over 4}-{\nu ^2} \bigr)-s(s+1)
\Bigr]\,{\1}, \qquad{Q_2} =\Bigl[ \bigl( {1\over 4}-{\nu ^2}
\bigl)s(s+1) \Bigr]\, {\1}.$$ Here, the only physical
representation in the sense of Poincar\'e limit is the scalar case
corresponding to $\Delta=(0,1)$ and also called conformally
coupled massless case.
\item  Discrete series $\Pi_{p,0}$ and $\Pi^{\pm}_{p,q}$:\quad
$\Delta=(p,q)$ with
\begin{eqnarray}
&&p=1,2,3,\dots  \quad {\rm and} \quad q=0 \quad {\rm or},\nonumber\\
&&p={1\over 2},1,{3\over 2},2,\dots \quad {\rm and} \quad
q=p,p-1,\dots,1\ {\rm or}\ {1\over 2}.\nonumber
\end{eqnarray}
In this case, the only physical representations in the sense of
Poincar\'e limit are those with $p=q=s$. They are called
the massless representations of the dS
group.
\end{enumerate}
Note that the substitution $q\rightarrow (1-q)$ does not alter the
eigenvalues; the representations with labels $\Delta=(p,q)$ and
$\Delta=(p,1-q)$ can be shown to be equivalent. Finally, we have
pictured some of these representations in terms of $p$ and $q$.
The symbols $\bigcirc$ and $\square$ stand for the discrete series
with semi-integer and integer values of $p$ respectively. The
complementary series is represented in the same frame by  bold
lines. The principal series is represented in the Re$(q)=1/2$
plane by dashed lines.
\begin{figure}[h]
\begin{center}
\begin{picture}(0,0)%
\includegraphics{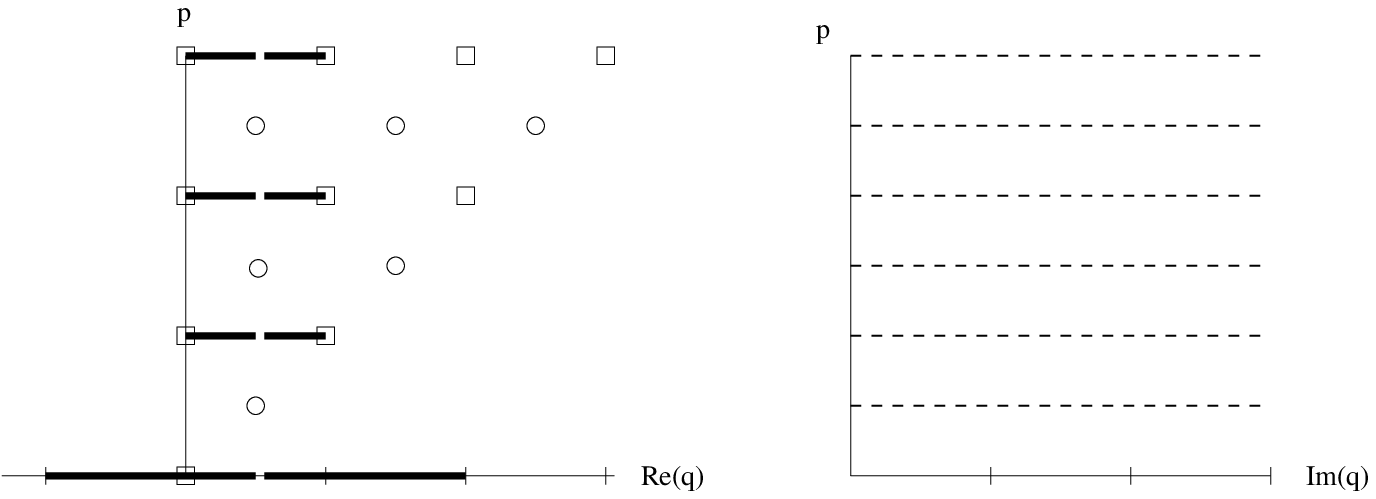}%
\end{picture}%
\setlength{\unitlength}{2210sp}%
\begingroup\makeatletter\ifx\SetFigFont\undefined
% extract first six characters in \fmtname
\def\x#1#2#3#4#5#6#7\relax{\def\x{#1#2#3#4#5#6}}%
\expandafter\x\fmtname xxxxxx\relax \def\y{splain}%
\ifx\x\y   % LaTeX or SliTeX?
\gdef\SetFigFont#1#2#3{%
  \ifnum #1<17\tiny\else \ifnum #1<20\small\else
  \ifnum #1<24\normalsize\else \ifnum #1<29\large\else
  \ifnum #1<34\Large\else \ifnum #1<41\LARGE\else
     \huge\fi\fi\fi\fi\fi\fi
  \csname #3\endcsname}%
\else \gdef\SetFigFont#1#2#3{\begingroup
  \count@#1\relax \ifnum 25<\count@\count@25\fi
  \def\x{\endgroup\@setsize\SetFigFont{#2pt}}%
  \expandafter\x
    \csname \romannumeral\the\count@ pt\expandafter\endcsname
    \csname @\romannumeral\the\count@ pt\endcsname
  \csname #3\endcsname}%
\fi \fi\endgroup
\begin{picture}(11187,4344)(214,-4261)
\put(5326,-4261){\makebox(0,0)[lb]{\smash{\SetFigFont{7}{8.4}{rm}$3$}}}
\put(1426,-2836){\makebox(0,0)[lb]{\smash{\SetFigFont{7}{8.4}{rm}$1$}}}
\put(1426,-1636){\makebox(0,0)[lb]{\smash{\SetFigFont{7}{8.4}{rm}$2$}}}
\put(1426,-436){\makebox(0,0)[lb]{\smash{\SetFigFont{7}{8.4}{rm}$3$}}}
\put(8551,-4261){\makebox(0,0)[lb]{\smash{\SetFigFont{7}{8.4}{rm}$1$}}}
\put(9751,-4261){\makebox(0,0)[lb]{\smash{\SetFigFont{7}{8.4}{rm}$2$}}}
\put(10951,-4261){\makebox(0,0)[lb]{\smash{\SetFigFont{7}{8.4}{rm}$3$}}}
\put(7051,-2836){\makebox(0,0)[lb]{\smash{\SetFigFont{7}{8.4}{rm}$1$}}}
\put(6976,-1636){\makebox(0,0)[lb]{\smash{\SetFigFont{7}{8.4}{rm}$2$}}}
\put(6976,-436){\makebox(0,0)[lb]{\smash{\SetFigFont{7}{8.4}{rm}$3$}}}
\put(376,-4261){\makebox(0,0)[lb]{\smash{\SetFigFont{7}{8.4}{rm}$-1$}}}
\put(1651,-4261){\makebox(0,0)[lb]{\smash{\SetFigFont{7}{8.4}{rm}$0$}}}
\put(2851,-4261){\makebox(0,0)[lb]{\smash{\SetFigFont{7}{8.4}{rm}$1$}}}
\put(4051,-4261){\makebox(0,0)[lb]{\smash{\SetFigFont{7}{8.4}{rm}$2$}}}
\end{picture}
\caption{SO$_{0}(1,4)$ unitary irreducible representation
diagrams.}
\end{center}
\end{figure}

%\begin{figure}[h]
%\begin{center}
%\input{rui1.pstex_t}
%\caption{SO$_{0}(1,4)$ unitary irreducible representation
%diagrams.}
%\end{center}
%\end{figure}

\section{Maximally symmetric bitensors in ambient space}
Following Allen and Jacobson in reference \cite{allen} we will
write the two-point functions in de Sitter space (maximally
symmetric) in terms of bitensors. These are functions of two
points $(x,x')$ which behave like tensors under coordinate
transformations at either point. The bitensors are called
maximally symmetric if they respect the de Sitter invariance.

As shown in reference \cite{allen}, any maximally symmetric
bitensor can be expressed as a sum of products of three  basic
tensors. The coefficients in this expansion are functions of the
geodesic distance $\mu(x,x')$, that is the distance along the
geodesic connecting the points $x$ and $x'$ (note that $\mu(x,x')$
can be defined by unique analytic extension also when no geodesic
connects $x$ and $x'$). In this sense,  these fundamental tensors
form a complete set.  They can be obtained by differentiating the
geodesic distance:
$$n_{a}=\nabla_{a}\mu(x,x'),\qquad n_{a'}=\nabla_{a'}\mu(x,x')$$
and the parallel propagator
$$ g_{ab'}=-c^{-1}(\z)\nabla_{a} n_{b'}+n_{a}n_{b'}\;.$$
The geodesic distance is implicitly defined \cite{brmo} for
$\z=-H^{2}x\cdot x'$ by
\begin{eqnarray*}
\z&=&\cosh(\mu H)\quad\mbox{for $x$ and $y$ timelike separated,}
\\
 \z&=&\cos(\mu H)\quad\mbox{for $x$ and $y$ spacelike separated such
 that}\quad \vert x\cdot x'\vert <H^{-2}.
\end{eqnarray*}
The basic bitensors  in ambient space notations are found through:
$$\bc\mu(x,x'),\qquad\bpc\mu(x,x'),\qquad \bc\bpc\mu(x,x'),$$
restricted to the hyperboloid by
$$T_{ab'}(x,x')=\ab T_{\alpha\beta'}\;.$$
For $\z=\cos(\mu H)$, one finds
$$n_{a}=\a\bar\partial_{\alpha}\mu(x,x')=\a{{H(\theta_{\alpha}\cdot
x')}\over{\sqrt{1-\z^2}}},\quad\qquad
n_{b}=\bb\bar\partial^{'}_{\beta'}\mu(x,x')=\bb{{H(\theta'_{\beta'}\cdot
x)}\over{\sqrt{1-\z^2}}}\;,$$ and
$$\nabla_{a}n_{b'}=\ab\theta^{\sigma}_{\alpha}\theta^{'\gamma'}_{\beta'}
\bar\partial_{\sigma}\bar\partial'_{\gamma'}\mu(x,x')=c(\z)\left[\z
n_{a}n_{b'}- \ab\theta_{\alpha}\cdot\theta'_{\beta'} \right] \,,$$
with $\displaystyle{c(\z)=-\frac{H}{\sqrt{1-\z^{2}}}}$. For
$\z=\cosh(\mu H)$, $ n_{a}$, $ n_{b'}$   are multiplied by $i$ and
$c(\z)$ becomes $-\frac{iH}{\sqrt{1-\z^{2}}}$. In both cases we
have
$$
\ab\theta_{\alpha}\cdot\theta'_{\beta'}
=g_{ab'}+(\z-1)n_{a}n_{b'}\;.
$$

\section{``Massive'' vector two-point function}
Given the important role played by the ``massive'' vector Wightman
two-point function in the construction of the spin-$2$ two-point
function we briefly present  here a derivation of it (for details
see Reference \cite{gata}). In addition we compare our two-point
function with the one given in Reference \cite{allen}. We consider
the ``massive'' vector Wightman two-point function which
corresponds to the principal series of representation of
SO$_{0}(1,4)$ and satisfies:
$$\left(Q_{1} -\langle Q_{1} \rangle\right){\cal
W}^{\nu}_{1\alpha\beta'}(x,x')=0,\qquad \mbox{where}\quad\langle
Q_{1} \rangle=\nu^2+{{1}\over{4}}\quad
\mbox{with}\quad\nu\in\setR\,.$$ This bivector is obtained as the
boundary value of the analytic bivector  two-point function
obtained with the modes (\ref{eq:vector}):
\begin{equation*}
W^\nu _{1\alpha \beta'}(z,z')= c_\nu^{2}\frac{\langle Q_{1}
\rangle}{\langle Q_{0} \rangle}\int_\gamma \sum_{\lambda}{\cal
E}^{\lambda}_{\alpha }(z,\xi) \,{\cal
E}^{*\lambda}_{\beta'}(z^{'*},\xi) (Hz\cdot\xi)^{\sigma}(H z'\cdot
\xi)^{\sigma^{*}}\,d\sigma_\gamma(\xi).
\end{equation*}
With the  help of Eq. (\ref{eq:prop1}) and the relation
$H^{2}D_{1}\left( Hz\cdot \xi\right)^{\sigma}=\bar\xi\left(
Hz\cdot \xi\right)^{\sigma}/\left( z\cdot \xi\right)$ it is easy
to expand the transverse bivector  in terms of the analytic scalar
two-point function $W_{0}(z,z')$:
\begin{equation}
W^{\nu}_{1}(z,z') =\frac{\langle Q_{0}\rangle }{\langle
Q_{1}\rangle}\left(-\,\theta_{\alpha}\cdot\theta'_{\alpha'}+
\frac{H^{2}\sigma(\theta\cdot z' )D'_{1}}{\langle Q_{0}\rangle }+
\frac{H^{2}\sigma^{*}(\theta'\cdot z) D_{1}}{\langle Q_{0}\rangle
}+ \frac{ H^2\z D_{1}D'_{1}}{\langle Q_{0}\rangle}\right)
W^{\nu}_{0}(z,z').\label{eq:vec}
\end{equation}
The analytic ``massive '' scalar two-point function is
\begin{equation*}
 W^\nu_{0}(z,z')= c_\nu^{2}\int_\gamma
(Hz\cdot\xi)^{\sigma}(H z'\cdot \xi
)^{\sigma^{*}}\,d\sigma_\gamma(\xi)\quad\mbox{with}\quad
c_{\nu}^{2}=H^2e^{+\pi\nu}\Gamma(-\sigma) \Gamma (-\sigma^*)/(2^5
\pi^4m^2)\,,
\end{equation*}
which satisfies:
$$\left(Q_{0} -\langle Q_{0} \rangle\right)W^{\nu}_{0}(z,z')=0,\qquad \mbox{where}\quad\langle Q_{0}
\rangle=\nu^2+{{9}\over{4}}\quad \mbox{with}\quad\nu\in\setR\,. $$
The choice of normalization corresponds to the Euclidean vacuum
and $W_0(z,z')$ can be written as a  hypergeometric function (see
\cite{brmo}):
$$ {\cal W}^\nu_{0}(z,z')= C_{\nu}\;
_{2}F_{1}\left(-\sigma,-\sigma^*;2;\frac{1+\z}{2}\right)= C_{\nu}
P_{\sigma}^{5}(-\z) \;\quad\mbox{with}\quad C_{\nu}=\frac{
H^2\Gamma(-\sigma)\; \Gamma (-\sigma^*)}{2^4 \pi^2m^2}.$$

In order to show that our vector two-point function  is the same
two-point function as the one given by Allen and Jacobson in
Reference \cite{allen}, we develop $W^{\nu}_{1}(z,z')$ using
essentially
$\bar\partial_{\alpha}\phi(\z)=-\left(\theta_{\alpha}\cdot
z'\right)H^{2}\dz \phi(\z).$ One finds
$$ {\cal W}^{\nu}_{1\alpha\beta'}(x,x')=\mbox{bv}\,W^{\nu}_{1}(z,z')=\theta_{\alpha}\cdot\theta'_{\beta'}U(\z)
+H^2{{(\theta'_{\beta'}\cdot z)\left(\theta_{\alpha}\cdot
z'\right)}\over{1-\z^2}}\;V(\z)\;,$$ with
\begin{equation}
U(\z)=-{{1}\over{\langle Q_1\rangle}} \left[Q_{0}+\z\dz\;\right]{
W^{\nu}_{0}(z,z')} \;,\quad V(\z)={{1}\over{\langle
Q_1\rangle}}\left[3\dz+\z^2\dz+\z Q_{0}\right]
W^{\nu}_{0}(z,z')\;,\nonumber
\end{equation}
where
$$Q_{0}=\left(1-\z^2\right)\ddz-4\z\dz,$$
is the second order differential operator deduced from the Casimir
operator expressed with the variable $\z$ in place of $(z,z')$.
The functions $U(\z)$ and $V(\z)$ satisfy the property
$$\z U(\z)+V(\z)={{3}\over{\langle Q_1\rangle}}\dz {W^{\nu}_{0}(z,z')}\equiv\Lambda(\z)\;,$$
with
$$\Lambda(\z)=3H^2\;\frac{ \Gamma(1-\sigma)\; \Gamma (1-\sigma^{*})}{2^6\langle Q_1\rangle \pi^2m^2}\;
_{2}F_{1}\left(1-\sigma,1-\sigma^{*};3;\frac{1+\z}{2}\right) \;,$$
which is solution of the equation
$$\left[Q_{0}-2\z\dz-6-\langle Q_{1}\rangle\right]\Lambda(\z)=0\;.$$

Finally, let us write the intrinsic expression of the two-point
function ${\cal W}^{\nu}_{1}(x,x')$  obtained as the boundary
value of $W^{\nu}_{1}(z,z')$. The intrinsic expression is:
$$Q_{ab'}\equiv\ab {\cal
W}^{\nu}_{1\alpha\beta'}(x,x')\;.$$ Since
$$
\ab\theta_{\alpha}\cdot\theta'_{\beta'}=g_{ab'}+
(\z-1)n_{a}n_{b'},\qquad \ab {{H^2(\theta'_{\beta'}\cdot
x)\left(\theta_{\alpha}\cdot x'\right)}\over{1-\z^2}}=n_{a}n_{b'},
$$
one gets
$$
Q_{ab}=g_{ab'}U(\z)+n_{a}\,n_{b'}\left(\Lambda(\z))-U(\z)\right),
$$
and in the case of SO$_{0}(4,1)$:
$$Q_{ab}=-g_{ab'}U(\z)-n_{a}\,n_{b'}\left(U(\z)-\Lambda(\z))\right).$$
This is the expression given by Allen and Jacobson in Ref.
\cite{gasp} and \cite{allen}.

\section{Another expression for the spin-$2$ Two-point function}
We present another form of the spin-$2$ two-point function, which
is useful for the proof of the locality condition. We begin with
the term $M(z,z')W_{1}^{\nu}(z,z')$:
\begin{equation*}
\frac{\langle Q_{0}\rangle+4}{\langle Q_{0}\rangle }\left[-\,{\cal
S }{\cal S' } \theta\cdot\theta'+\frac{H^{2}{\cal S }(\theta\cdot
z')D'_{2}}{\sigma^{*}-1}+\frac{H^{2}{\cal S' }(\theta'\cdot
z)D_{2}}{\sigma-1}+\frac{\z H^{2}D_{2}D'_{2}}{
(\sigma-1)(\sigma^{*}-1)}\right]W_{1}^{\nu}(z,z')\,.
\end{equation*}
We rewrite this equation using the relations
\begin{eqnarray*}
&&H^{2}{\cal S }(\theta\cdot z')D'_{2} W_{1}^{\nu}=-\,{\cal S
}{\cal S' } \theta\cdot\theta'  W_{1}^{\nu}+2\theta' {\cal
S}\theta\cdot
W_{1}^{\nu}+\frac{1}{2}D_2D_2'W_{2}\,,\nonumber \\
&&H^{2}{\cal S }'(\theta'\cdot z)D_{2} W_{1}^{\nu}=-\,{\cal S
}{\cal S' } \theta\cdot\theta'  W_{1}^{\nu}+2\theta {\cal
S}'\theta'\cdot W_{1}^{\nu}+\frac{1}{2}D_2D_2'W_{2}\,, \nonumber\\
&&H^{2}\z D_2D'_{2} W_{1}^{\nu}=-\,{\cal S }{\cal S' }
\theta\cdot\theta' W_{1}^{\nu}+2\theta {\cal S}'\theta'\cdot
 W_{1}^{\nu}+2\theta' {\cal S}\theta\cdot
W_{1}^{\nu} +D_2D_2'\left(W_{2}+H^{2}\z W_{1}^{\nu}\right)\,,
\end{eqnarray*}
where $D_2D_2' W_{2}=2H^{2}D_{2}{\cal S}'\left(\theta'\cdot z
\right)W_1 -4\theta{\cal S}'\theta'\cdot W_1$. This is obtained by
simple calculation of
$$
\left(Q_{2}-\langle Q_2\rangle\right) \left({\cal S }{\cal S' }
\theta\cdot\theta' W_{1}^{\nu}+D_2 D_2'W_3\right)=0 \,,$$ with the
help of Eq. (\ref{eq:prop4}) and where we have written
$W_{2}=\left(Q_{1}-\langle Q_2\rangle\right)W_3$. One gets:
\begin{equation*}
M(z,z')W_{1}^{\nu}(z,z')= -\,{\cal S }{\cal S' }
\theta\cdot\theta'  W_{1}^{\nu}+\frac{2\theta\sigma^{*}{\cal
S}'\theta'\cdot
 W_{1}^{\nu}}{\langle
Q_0 \rangle} +\frac{2\theta' \sigma{\cal S}\theta\cdot
W_{1}^{\nu}}{\langle Q_0 \rangle} +D_2D_2'\left(\frac{H^{2}\z
W_{1}^{\nu}}{\langle Q_0 \rangle}-\frac{3W_{2}}{2\langle Q_0
\rangle}\right)\,.
\end{equation*}

Now, given that
\begin{equation*}\label{eq:prop2} {\cal
S}\theta\cdot W^{\nu}_{1}=\frac{2}{3}\left[
\theta+\frac{H^{2}D_{2}D_{1}}{2\langle Q_{0}\rangle}\right](
W^{\nu}_{1} )' \quad\mbox{and}\quad{\cal S}'\theta'\cdot
W^{\nu}_{1}=\frac{2}{3}\left[
\theta'+\frac{H^{2}D'_{2}D'_{1}}{2\langle
Q_{0}\rangle}\right](W^{\nu}_{1} )'\,,
\end{equation*}
where $(W_{1}^{\nu})'$ is the trace of the vector two-point
function given by
$$(W^{\nu}_{1})'=\eta\cdot\cdot  W^{\nu}_{1}=3U(\z)+\z\Lambda(\z)
=-3\frac{\langle Q_{0}\rangle}{\langle
Q_{1}\rangle}\;W^{\nu}_{0}(z,z')\,.
$$
We find the following form  for the spin-2 two-point function :

\begin{eqnarray*}
W^{\nu}(z,z')&=&M(z,z')W_{1}^{\nu}(z,z')+N(z,z')W_{0}^{\nu}(z,z')\nonumber
\\ &=& q\left[\theta \theta' (W_{1}^{\nu})'-\frac{3}{2}\theta{\cal
S}'\theta'\cdot W^{\nu}_{1}- \frac{3}{2}\theta'{\cal S}\theta\cdot
W^{\nu}_{1}\right] -\,{\cal S }{\cal S' } \theta\cdot\theta'
W_{1}^{\nu} +D_2D_2'W_{4}\,,
\end{eqnarray*}
where $q=-\frac{4}{9}\left({\langle
Q_{0}\rangle-9}\right)/{\langle Q_{0}\rangle}$ and
\begin{equation*}
D_{2}D'_{2} W_{4}=\frac{6\theta{\cal S }'\theta'\cdot
W^{\nu}_{1}}{\langle Q_{0}\rangle}-\frac{3H^{2}D_{2}{\cal S
}'(\theta'\cdot z)W^{\nu}_{1}}{\langle Q_{0}\rangle}
+D_{2}D'_{2}\left(\frac{H^{2}\z   W^{\nu}_{1}}{\langle
Q_{0}\rangle}+\frac{H^{4}D_{1}D'_{1}( W^{\nu}_{1})'}{9\langle
Q_{0}\rangle^{2}}\right)\,.
\end{equation*}
The two-point function can be rewritten as
$$W^{\nu}(z,z')=D(z,z')W_{0}^{\nu}(z,z')\,,$$
where the differential operator $D(z,z')$ obviously satisfies
$D^{*}(z^{*},z^{'*})=D(z,z')\,$. This property serves to prove the
locality condition.
\end{appendix}

\end{document}